\documentclass[epj,referee]{svjour}

\usepackage{latexsym}

\usepackage{afterpage}
\usepackage{cite}

\usepackage{amsfonts}
\usepackage{amssymb}
\usepackage{subfigure}
\usepackage{graphicx}

\usepackage{textcomp}
\usepackage{gensymb}
\usepackage[english]{babel}
\usepackage[T1]{fontenc}
\usepackage{setspace}
\usepackage{color}
\usepackage{siunitx}
\usepackage{isotope}
\usepackage{here}
\usepackage{afterpage}
 
\begin{document}

\title{Neutron optics of the 
PSI ultracold neutron source:
characterization and simulation
}

\author{
G.~Bison\inst{1},
B.~Blau\inst{1},
M.~Daum\inst{1},
L.~G\"oltl\inst{1},
R.~Henneck\inst{1},
K.~Kirch\inst{1,2},
B.~Lauss\inst{1}$^,$\thanks{Corresponding author, tel. +41(0)56 3104647, bernhard.lauss@psi.ch, ucn.web.psi.ch}
D.~Ries\inst{1,2}$^,$\thanks{present address: Institute for Nuclear Chemistry, Johannes-Gutenberg-University, Mainz, Germany},
P.~Schmidt-Wellenburg\inst{1},
G.~Zsigmond\inst{1}$^,$\thanks{Corresponding author, tel. +41(0)56 3103569, geza.zsigmond@psi.ch, ucn.web.psi.ch}
}

\institute{$^1$
Paul Scherrer Institut, CH-5232 Villigen-PSI, Switzerland \\
$^2$
ETH Z\"urich, Z\"urich, Switzerland \\
}


\date{\today}

\abstract{
The ultracold neutron (UCN) source at the Paul Scherrer Institute
serves mainly experiments in fundamental physics.
High UCN intensities are the key for progress and success
in such experiments.
A detailed understanding of all source parameters is required
for future improvements.
Here we present the UCN source components,
elements of the neutron optics,
the characterization of important related parameters
like emptying times, storage times or transmission probabilities of UCN
which are ultimately defining the UCN intensity delivered at the beamports.
We also introduce
a detailed simulation model
of the PSI UCN source,
used to analyze the measurements
and to extract surface parameters.
}


\authorrunning{G.~Bison et al.}
\titlerunning{UCN optics}

\maketitle

\onecolumn



\section{Introduction}

Neutrons with velocities below about 8\,m/s and hence 
with kinetic energies below about 300\,neV,
equivalent to temperatures below 3\,mK 
are termed ultracold neutrons (UCN).
Such neutrons have unique properties~\cite{Golub1991,Ignatovich1990}.
Most prominently, 
UCNs are reflected under any angle of incidence from suitable materials
like steel, beryllium, nickel, diamond-like carbon (DLC) or 
nickel-molybdenum (NiMo).
Hence, UCNs can be stored in bottles
made from or coated with these materials, 
where they can be observed for hundreds of seconds, 
in principle only limited by the neutron lifetime.
Moreover, UCN can also be guided to remote areas over tens of meters 
and behave more comparable to an extremely dilute ideal gas
than to a particle beam.
The specific description of UCN transport includes basic
transport processes and 
includes also magnetic interactions, various loss mechanisms and
gravity.
All together can be summarized with the term UCN optics~\cite{Ignatovich2010}.

The distinct behavior makes UCN an ideal tool to study the 
intrinsic properties of the neutron
and also to search for physics beyond the Standard Model of particle physics (BSM),
most prominently represented by experiments searching for a 
possible permanent electric dipole moment of the neutron 
(nEDM)~\cite{Baker2006,Baker2011,SerebrovEDM2015,Pendlebury2015,Abel2017,Kolarkar2010,Martin2013}
or the precise determination of the neutron lifetime~\cite{Wie2011,Yue2013,Serebrov2018,Pattie2018}.

Such precision measurements are presently limited by neutron counting statistics, 
hence there are worldwide efforts to build new 
UCN sources with higher intensities~\cite{Kirch2010,Bison2017}.
A new high-intensity UCN source was designed 
and constructed at the Paul Scherrer Institute (PSI), Villigen, Switzerland, 
and started operation in 2011.
Specific details and efforts were already reported in 
Refs.~\cite{Anghel2009,Lauss2011,Lauss2012,Lauss2014,Becker2015,Blau2016,Goeltl2012,Ries2016,Bison2017}.
Here we compare source characterization measurements
to detailed
Monte-Carlo (MC) simulations 
using the simulation tool 
MCUCN~\cite{Zsigmond2018}
developed at PSI.
This provides input to a better 
understanding of the UCN source performance.


Further increase of UCN intensities requires an
excellent understanding of all components of the UCN source.
Scope of the presented Monte-Carlo analysis is the extraction
of parameters as defined in established UCN physics in order 
to characterize the quality of the optical components, 
and to identify possible improvements. 
For this purpose, all processes which could lead to a loss of UCN have to be 
identified, monitored and eliminated or reduced as much as possible.
For instance, losses due to frost formation on the solid deuterium (sD$_2$) surface
have been only recently identified together with a way to 
recover performance~\cite{Anghel2018}.


This paper is organized in the following way.
In Sec.~\ref{sec:UCN-source} we give an introduction to all setup 
items of the 
UCN source which are relevant to its neutron optics performance.
Section~\ref{sec:simulation} provides an introduction to our simulation 
model describing the UCN transport in the source and its guides.
In Sec.~\ref{sec:characterization} we describe measurements which provide a
characterization of different important neutron optics parameters of the operating UCN source.
Simulations used for the analysis of characterizing parameters are described
in Sec.~\ref{sec:simulation-analysis},
followed by short 
conclusions in Sec.~\ref{sec:summary}.


\section{UCN source setup}   
\label{sec:UCN-source}

\begin{figure}[htb]
\begin{center}
\resizebox{0.5\textwidth}{!}{\includegraphics{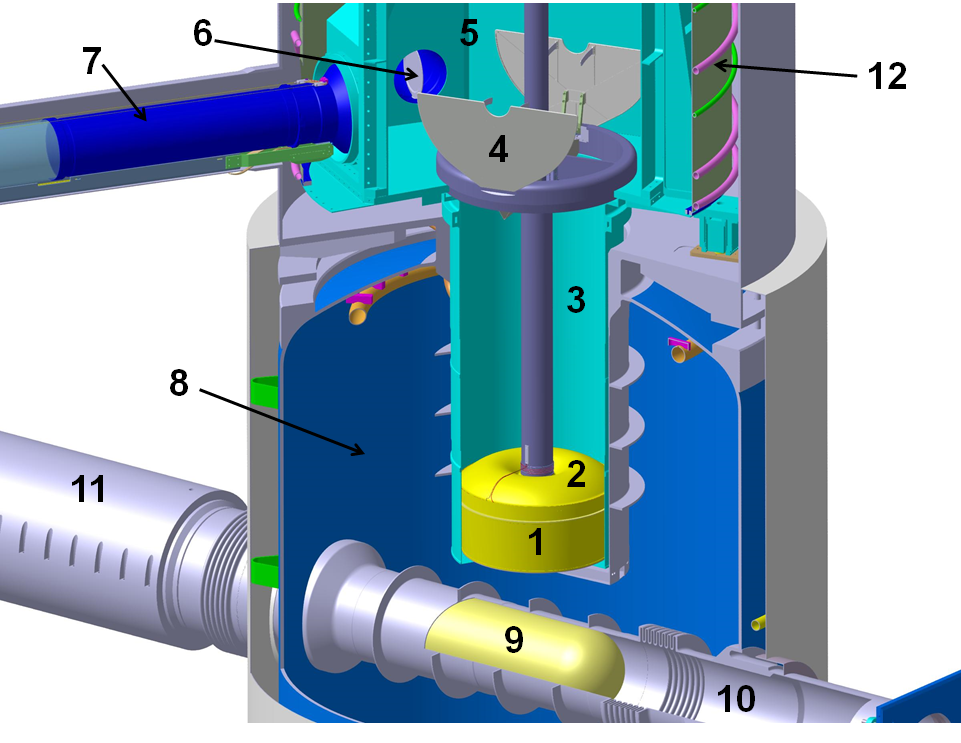}}
\caption [Sketch of the central section of the UCN source] {
CAD image of the lower section of the UCN tank with indicated
parts relevant to UCN production and transport (see text).
1 - D$_2$ moderator vessel, 
2 - lid,
3 - vertical guide,
4 - storage vessel flapper valve,
5 - storage vessel,
6 - UCN guide shutter,
7 - UCN guide section,
8 - heavy water moderator tank,
9 - spallation target,
10 - proton beam tube,
11 - target shielding,
12 - thermal shield.
}
\label{fig:UCN-source}
\end{center}
\end{figure}

\begin{figure*}[htb]
\begin{center}
\resizebox{0.85\textwidth}{!}{\includegraphics{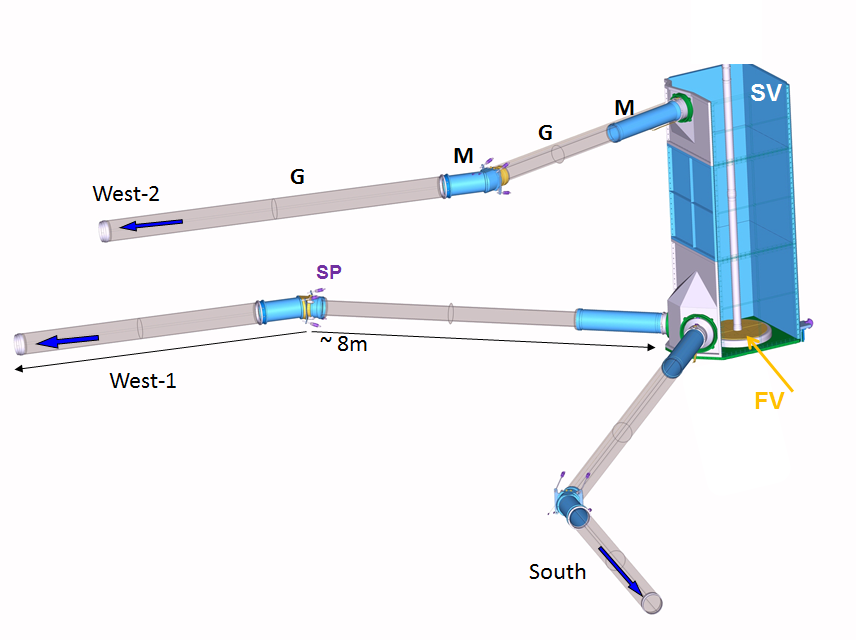}
}
\caption [UCN source with guides] {
View of the three UCN guides 
and UCN storage vessel (SV) with closed flapper valve (FV).
UCN guides made from glass tubes (G) are shown in light gray,
metal tubes (M) are shown in darker blue,
as indicated on guide West-1.
Mechanical springs (SP) keep the positioning.
The length of the UCN guide West-1 from 
SV to area West is approximately 8\,m.
}
\label{fig:UCN-source-guides}
\end{center}
\end{figure*}

Overviews of the main parts of the UCN source relevant to this work
are shown in 
Figures~\ref{fig:UCN-source} and
\ref{fig:UCN-source-guides}.
%
Many important subsystems 
devoted to cooling and cryo-operation~\cite{Anghel2008}, 
vacuum, 
gas handling,
heavy water system, 
spallation target~\cite{Meer2004,Wohlmuther2006}
and the proton beam line~\cite{Anghel2009,Anicic2005}
are omitted here.

The source operation is as follows:
The full proton beam of 590\,MeV and up to 2.4\,mA
is directed 
onto a spallation target~\cite{Wohlmuther2006}
with a pulse length of up to 8\,s.
About 8 free neutrons per incident proton
are produced in the target made of lead~\cite{Atchison2005d}.
These neutrons are 
thermalized in the surrounding heavy water (D$_2$O) 
at room temperature (about 30$\degree$C during operation). 
Thermal neutrons are further moderated and down-scattered into the 
cold and ultra-cold range inside solid 
deuterium (sD$_2$),
%
kept at a temperature of 5\,K inside the moderator vessel
located about 40\,cm above the spallation target.
%
%
UCN production in sD$_2$ has been investigated over 
many years~\cite{
Golub1977,
Serebrov1995b,
Serebrov1997,
Serebrov2000,
Serebrov2001b,
Morris2002,
Saunders2004,
Bodek2004,
Atchison2005,
Atchison2005a,
Atchison2005b,
Atchison2007,
Atchison2011
}.

UCN leaving the sD$_2$ gain kinetic energy 
by the material optical potential of 102\,neV~\cite{Daum2008}.
The UCN 
traverse a 0.5\,mm thin AlMg3 lid
separating the sD$_2$ from the UCN storage vacuum.
They are further guided upwards through a 
flapper valve into the storage vessel.
At the end of the proton beam pulse 
the valve is closed 
and reopened about 20\,s before the next pulse.

On the bottom of the storage vessel two 
UCN guides, called ``West1'' and ``South'', 
connect the source to the
corresponding beamports and 
experimental areas, West and South.
Figure~\ref{fig:UCN-source-guides}
shows a CAD image of this guide arrangement.
The upper guide, ``West-2'',
with lower intensity is used for test measurements.
Neutron shutters are positioned at the 
storage vessel exits and allow to close the
vessel for UCN storage.

\begin{figure}[htb]
\begin{center}
\resizebox{0.50\textwidth}{!}{\includegraphics{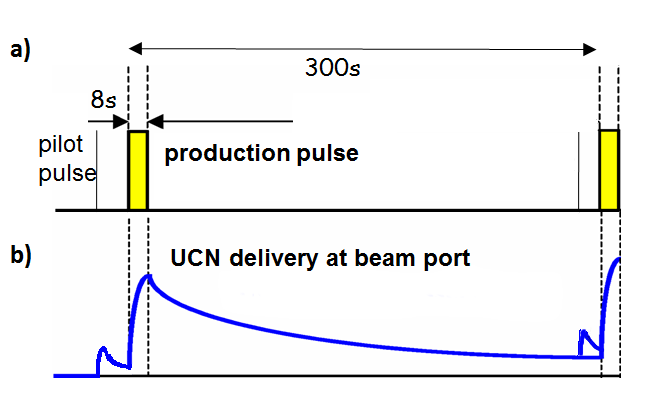}
}
\caption [Proton pulse sequence] {
A typical pulse sequence during the operation of 
the UCN source. 
a) 
Sketch of the proton beam pulse timing towards the UCN spallation target.
A 5\,ms long pilot pulse is used to monitor
beam centering and size. 
If approved, the main proton beam pulse
with a maximum duration of 8\,s 
is delivered 10\,s later. 
A kick signal coincident with the pulse start 
triggers the flapper valve (No.4 in Fig.~\ref{fig:UCN-source}) 
closing sequence.
The sequence can for example be every 300\,s.\newline
b) 
Observed time distribution of UCN counts measured at the West-1 beamport.
}
\label{fig:pulse-sequence}
\end{center}
\end{figure}

Proton beam operation for UCN production consists of 
single pulses
of up to 8\,s duration
with a typical repetition period of 300\,s. 
The maximum pulse length results from the constraint 
to keep the temperature rise 
inside the
sD$_2$ within a few Kelvin during the pulse.
A kicker magnet was specially designed to allow for a fast 
kick of the full proton intensity to the 
UCN proton beamline with a rise and fall time 
of 2.5\,ms~\cite{Anicic2005}.
A typical pulse sequence 
during standard operation is sketched in Fig.~\ref{fig:pulse-sequence}
indicating the duration and repetition of the 
proton pulse and the UCN intensity observed at a beamport.

Below we give a detailed description of key components for UCN transport, 
i.e.\ items with surfaces in contact with UCN, 
which are also included in the simulation model.

\subsection{D$_2$ vessel and lid}

\begin{figure}[htb]
\begin{center}
\begin{subfigure}[]
{\includegraphics[width=0.30\textwidth]{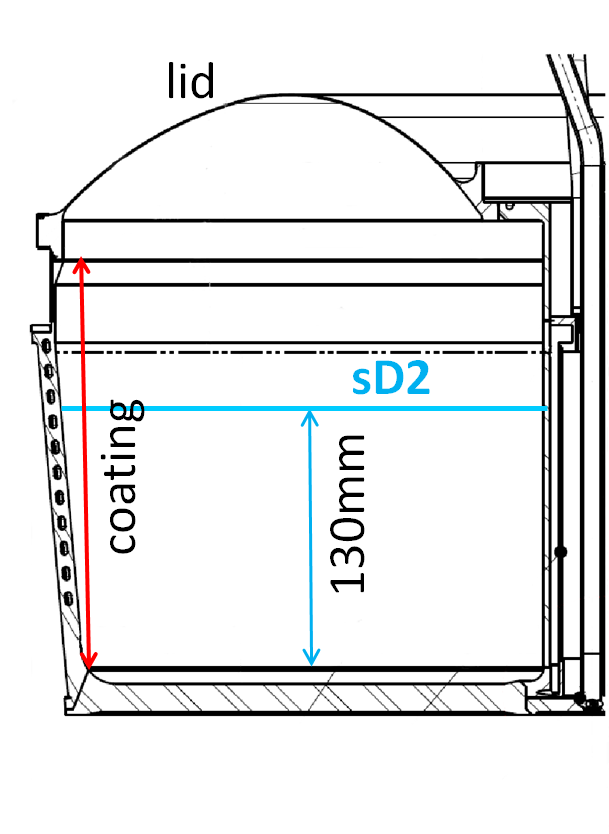}}
\end{subfigure}
\begin{subfigure}[]
{\includegraphics[width=0.35\textwidth]{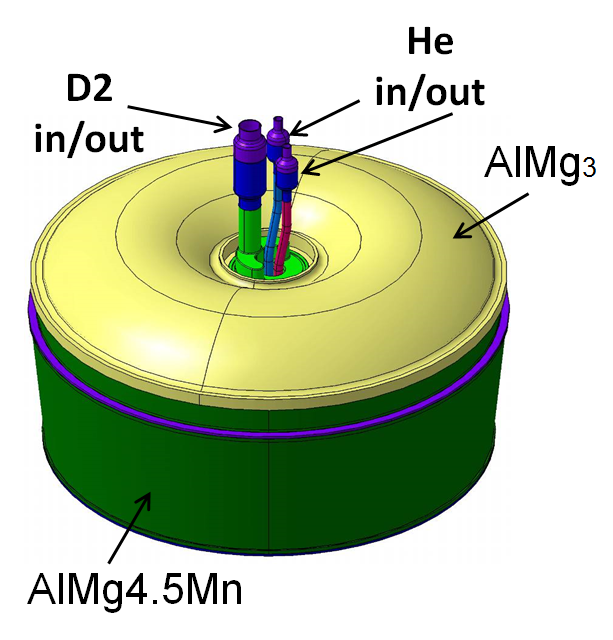}}
\end{subfigure}
\caption [D$_2$ vessel] {
a) Cut view drawing of half of the moderator vessel with indicated 
height of coating and sD$_2$.
The dark circles on the outside wall indicate the cooling channels.
\newline
b) CAD image of the moderator vessel.
The D$_2$ filling tube is indicated. 
The two smaller tubes are used for
supercritical He coolant in- and out-let.
}
\label{fig:D2-vessel}
\end{center}
\end{figure}

\begin{figure}[htb]
\begin{center}
\resizebox{0.45\textwidth}{!}{\includegraphics{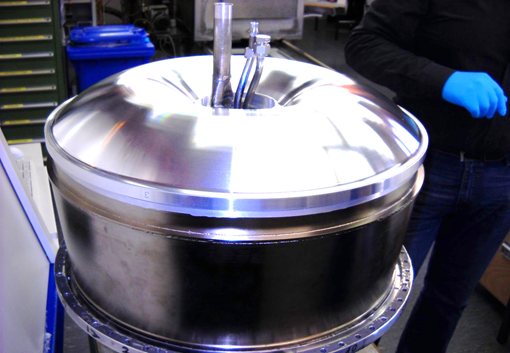}
}
\caption [D$_2$ vessel photo] {
Photo of the sD$_2$ moderator vessel after assembly.
}
\label{fig:D2-vessel-pic}
\end{center}
\end{figure}

The D$_2$ moderator vessel is a sophisticated double-walled container
with a volume of 44\,l,
engineered for cooling with super-critical helium at 4.9\,K.
A CAD image is 
shown in Fig.~\ref{fig:D2-vessel} together with a cut view.
The cooling channels were wire-eroded into the bottom plate and the side walls
(visible in the cut view)
before all parts were welded together.
A photo of the final vessel is shown in Fig.~\ref{fig:D2-vessel-pic} .

Relevant for UCN transport is the coating on the inside of the vessel.
The vessel with the exception of the top lid
was galvanically coated
with about 70\,$\mu$m of nickel.
In a second step, using sputter-coating\footnote{S-DH GmbH Heidelberg
Hans-Bunte-Str.8-10,
69123 Heidelberg, Germany, http://www.s-dh.de},
a thin layer of about 400\,nm
of a nickel-molybdenum (NiMo) alloy with a 85/15 weight 
percent mixture
was added on top.

The vessel lid with a thickness of 0.5\,mm
was machined from a single forged piece of AlMg3.
A tolerance in thickness of less than $\pm$50\,$\mu$m
was maintained during production.
This thin lid is necessary as the thickness of the material 
has a strong influence on UCN transmission~\cite{Atchison2009}.
In Fig.~\ref{fig:D2-vessel-pic} the toroidally-shaped lid is clearly visible.

\subsection{Vertical UCN guide}

\begin{figure}[h]
\begin{center}
\begin{subfigure}[]
{\includegraphics[width=0.3\textwidth]{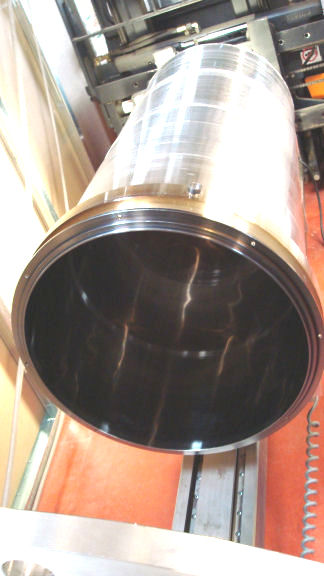}}
\end{subfigure}
\begin{subfigure}[]
{\includegraphics[width=0.4\textwidth]{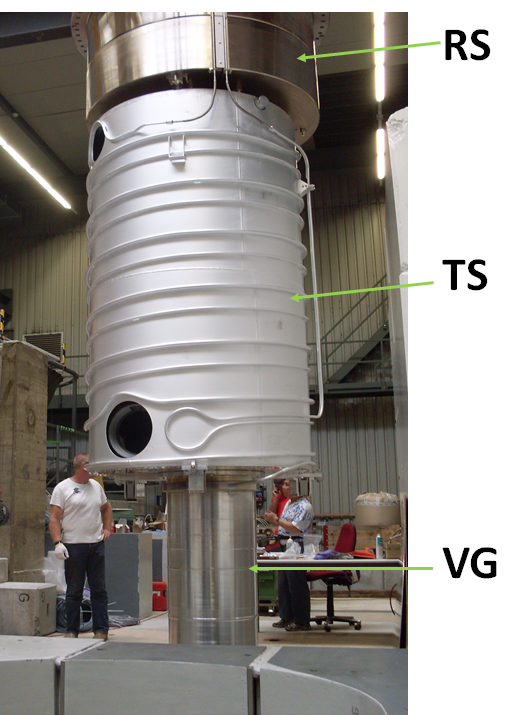}}
\end{subfigure}
\caption [Vertical guide] {
a) View of the vertical guide after coating 
with a highly reflective surface inside.\newline
b) Final assembly with vertical guide (VG) on the bottom,
thermal shield (TS) for the storage vessel,
and stainless-steel radiation shield (RS). 
The two visible big openings are for the UCN ports,
South (bottom) and West-2 (top).
}
\label{fig:vertical-guide}
\end{center}
\end{figure}

The vertical UCN guide is indicated as (3) in Fig.~\ref{fig:UCN-source}
and displayed in Fig.~\ref{fig:vertical-guide}a. 
The sD$_2$ vessel is located inside, at the bottom of the vertical guide
which serves also as a cryo-shield.
During operation the vertical guide is cooled to 
about 80\,K.
This guide bridges about one meter of height between moderator lid and 
UCN storage vessel.
This vertical rise
decreases the UCN kinetic energies by roughly the same amount as they gain 
via the energy boost when exiting the sD$_2$.

\begin{figure}[htb]
\begin{center}
\resizebox{0.80\textwidth}{!}{\includegraphics{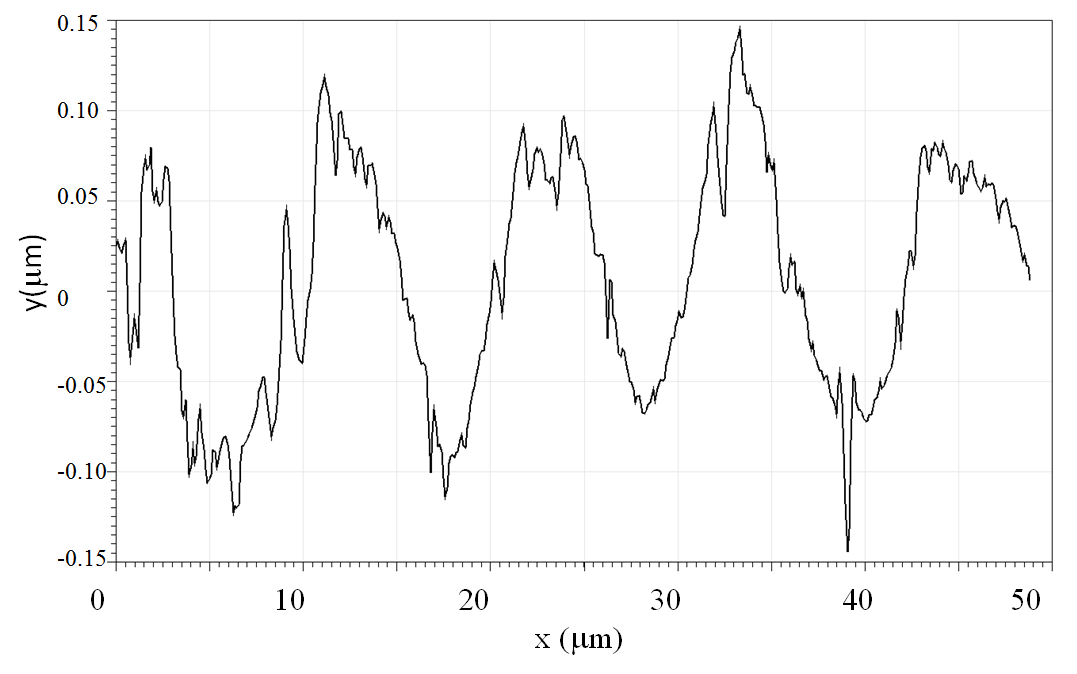}
}
\caption [AFM vertical guide] {
Profile view of an AFM scan before coating 
of the surface of the vertical guide
over a length of 50$\mu$m
showing a groove depth of $\pm$100\,nm.
Due to the softness of the pure aluminum 
the surface showed 
a typical structure coming from the 
diamond-lathe machining.
}
\label{fig:AFM-vertical-guide}
\end{center}
\end{figure}

The guide cylinder
is made from ultra-pure aluminum.
Four 25 cm-thick cylindrical pieces of aluminum were
diamond-lathed on the inside to
a surface roughness better than 100\,nm,
then machined with wall thicknesses of 2, 4 and 6\,mm, 
and then welded together.
Finally, the inside was sputter-coated with 400\,nm of 
nickel-molybdenum.
The surface roughness of the diamond-lathed surface
was determined using 
atomic force microscopy (AFM). 
The scan of one cylindrical piece
is shown in Fig.~\ref{fig:AFM-vertical-guide}.
The grooves from lathe machining 
are visible 
with a groove depth of about $\pm$100\,nm,
which can be 
determined in the profile view.

\subsection{UCN storage vessel}

\begin{figure}[h]
\begin{center}
\begin{subfigure}[]
{\includegraphics[width=0.45\textwidth]{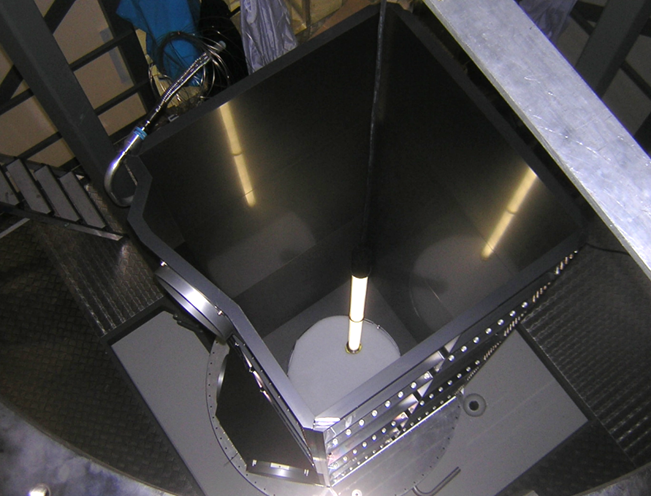}}
\end{subfigure}
\begin{subfigure}[]
{\includegraphics[width=0.45\textwidth]{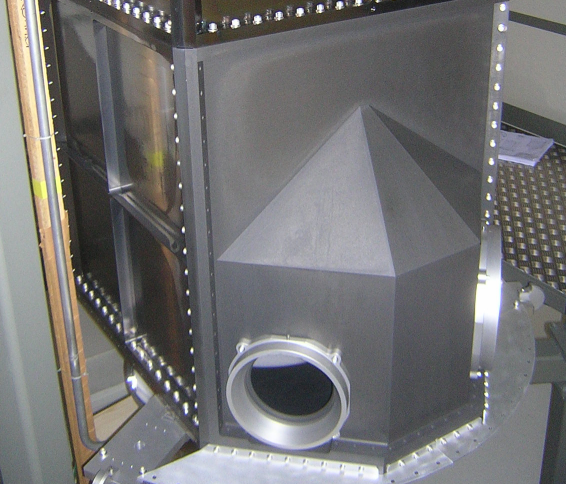}}
\end{subfigure}
\caption [Storage vessel] {
a) View inside the assembled UCN storage vessel from the top.
On the left the exit for guide West-2 is visible. 
On the bottom the circular hole connecting to the vertical guide
is visible.
The reflection of the lamp demonstrates the surface smoothness
and dark gray color of the diamond-like carbon coating.
b) Outside of the guide exit section after coating.
The large openings are the connection ports for the UCN guides,
West-1 (left) and Southt (right).
}
\label{fig:storage-vessel}
\end{center}
\end{figure}

The UCN storage vessel, indicated as (SV) in Fig.~\ref{fig:UCN-source-guides}
has a volume of about 1.58\,m$^3$ and an inside surface of 95'500\,cm$^2$.
During operation it is cooled to around 80\,K.
It is made from machined aluminum plates which were joined
carefully together to prevent gaps 
between the plates
as well as shape modification of the assembled vessel during cool down.
Depending on the size such gaps can cause large UCN losses.
The total area of gaps
of the assembled storage vessel was therefore minimized.
A rough mechanical measurement at room temperature
resulted in a gap fraction of 
about 5$\times$10$^{-4}$ of the total surface.
Two photos of the storage vessel are shown in Fig.~\ref{fig:storage-vessel}.

The inside surfaces of the storage vessel open to UCN are
all machined to a roughness of about 400\,nm.
The plates were galvanically coated with nickel 
(thickness about 60\,$\mu$m)
and finally 
coated with diamond-like carbon 
(DLC)\footnote{Fraunhofer-Institut f\"ur Werkstoff- und Strahltechnik,
Fraunhofer Projektgruppe at the Dortmunder Oberfl\"achenzentrum,
Eberhardstrasse 12, 44120 Dortmund, Germany, www.iws.fraunhofer.de}
which was found to have excellent UCN 
storage properties~\cite{Atchison2005c,Atchison2006,Atchison2006b,Atchison2007a,Atchison2007b,Atchison2008}.
The thickness of the DLC coating varied between 1 and 2\,$\mu$m depending on  
position during the coating process.
The material optical potential of a sample plate coated with DLC 
in the same process as the storage vessel plates
was measured at a cold-neutron reflectometer~\cite{Bodek2008}.
%
A neutron optical potential 
between of 235$\pm10$\,neV
was determined.

%
%

\begin{figure}[htb]
\begin{center}
\resizebox{0.5\textwidth}{!}{\includegraphics{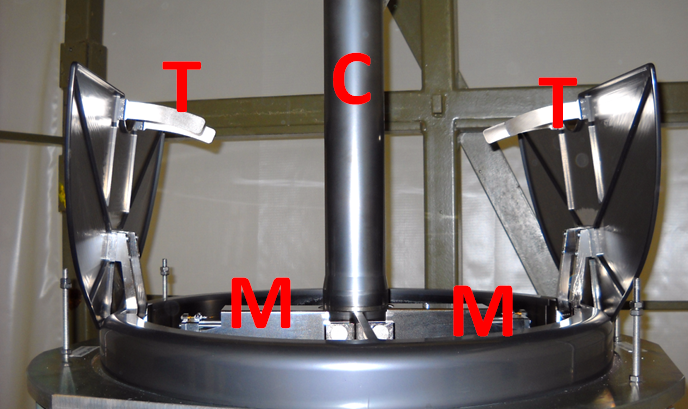}
}
\caption [Flapper valve] {
Photo of the open flapper valve in the test setup.
The visible `teeth' (T) are made from pure aluminum 
and serve in combination with magnets (M)
as eddy-current brakes to
smoothen the flap closing.
(C) indicates the central tube.
}
\label{fig:flapper-valve}
\end{center}
\end{figure}

The central tube
inside the storage vessel
serves as supply tube for cyrogenic liquids 
to the moderator vessel and is shown
in Fig.~\ref{fig:flapper-valve}. 
It is made of Al and
coated with DLC on the outside, 
identically as the vessel plates.
%
%
The hole on the bottom of the vessel can be closed 
by a valve made of
two semi-circular flaps
with 57\,cm diameter,
indicated as (4) in Fig.~\ref{fig:UCN-source}.
A photo of the flapper valve is
shown in Fig.~\ref{fig:flapper-valve}.
Machined from AlMgSi, the flaps were also
galvanically coated with nickel and subsequently coated with DLC.

\begin{figure}[htb]
\begin{center}
\resizebox{0.80\textwidth}{!}{\includegraphics{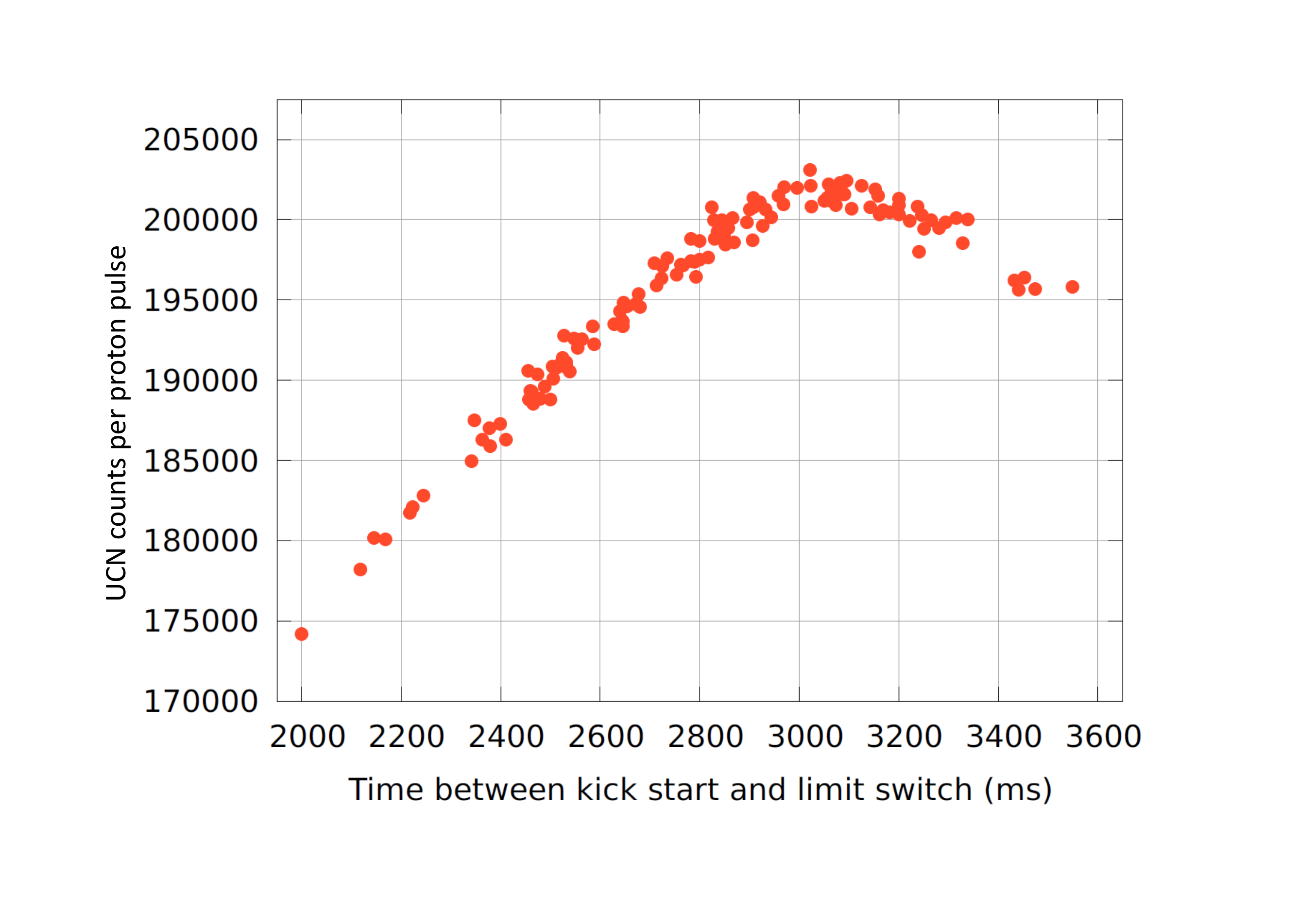}
}
\caption [Timing optimization] {
Correlation plot of the measured UCN counts 
at the West-2 beam port
and 
the measured time difference between the kick signal and 
the time when the first flap of the 
flapper valve starts closing.
For the measurement, the delay of the command to close 
the flapper valves has been varied. 
The intensity maximum is selected to be the
optimized closing time.
}
\label{fig:flapper-timing}
\end{center}
\end{figure}

Closing of the flapper valve is triggered by a signal 
synchronized with the proton beam kick
and done with a pneumatic actuator using helium.
The closing time is around two seconds.
Opening of the flaps takes about 10\,s 
because the high impedance of the long helium feeding line results 
in a slow pressure build-up in the actuator cylinder.
The slow opening does not significantly influence the UCN performance.
Timing between the flap-closing signal and the beam signal was optimized 
and shows a $\pm$100\,ms flat maximum as seen
in Fig.~\ref{fig:flapper-timing} 
displaying
the UCN count rate at beamport West-2 
as function of 
the closing time of the flapper valve.
In practice this is the time difference
between the kick signal 
coincident with the rising flank of the proton beam pulse 
(see Fig.\ref{fig:pulse-sequence})
and the time when the flapper valve starts closing.

\subsection{UCN guides}

Three UCN guides connect the storage vessel to the beamports 
in the experimental areas, see Fig.~\ref{fig:UCN-source-guides}.
The guide quality, mainly defined by 
low surface roughness and 
uniform coating with suited materials, 
is of paramount importance in order 
to minimize UCN losses
on the long path to experiments.
The length of the individual guide sections to the beamport West-1 is 7.090\,m,
to West-2 is 8.618\,m,
and to South is 7.049\,m.
The diameter was optimized by MC simulations to find a 
compromise between better transmission (large diameter) 
and less UCN density dilution (small diameter).

The straight guide sections were made from 
Duran\textsuperscript{TM} glass 
tubes\footnote{
SCHOTT AG, Hattenbergstr. 10, 55122 Mainz, Germany}
with a surface roughness
better than 1\,nm and therefore an expected high UCN 
transmission~\cite{Atchison2010}.
The inside diameter of the tube is 180\,mm,  
the wall thickness is 5\,mm.

Two sections per guide were made from stainless steel:
1) A bent was necessary for radiation protection
to prevent a direct line of sight between experimental 
area and storage vessel.
2) Stainless steel was chosen as it
increases the radiation hardness
of the first meter guide docking directly onto the storage vessel
and containing the neutron guide shutter.

\begin{figure}[htb]
\begin{center}
\resizebox{0.8\textwidth}{!}{\includegraphics{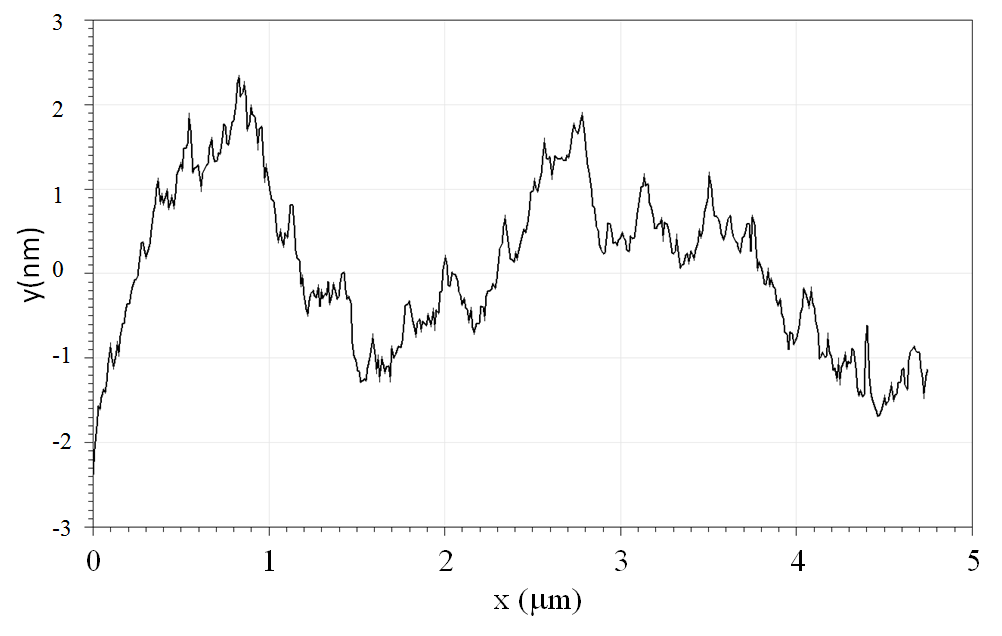}
}
\caption [AFM of glass] {
Display of an AFM measurement of the smooth glass surface inside a
UCN guide over an area of 5\,$\mu$m side-length.}
\label{fig:AFM-Duran}
\end{center}
\end{figure}

All metal tubes were honed on the inside and then hand-polished
to a surface roughness below 10\,nm.
All surfaces exposed to UCN
were characterized using atomic-force microscopy as 
shown for a typical glass surface 
in Fig.~\ref{fig:AFM-Duran}. 
The profile view shows a smoothness for a glass tube to be below 2\,nm.

\begin{figure}[htb]
\begin{center}
\resizebox{0.5\textwidth}{!}{\includegraphics{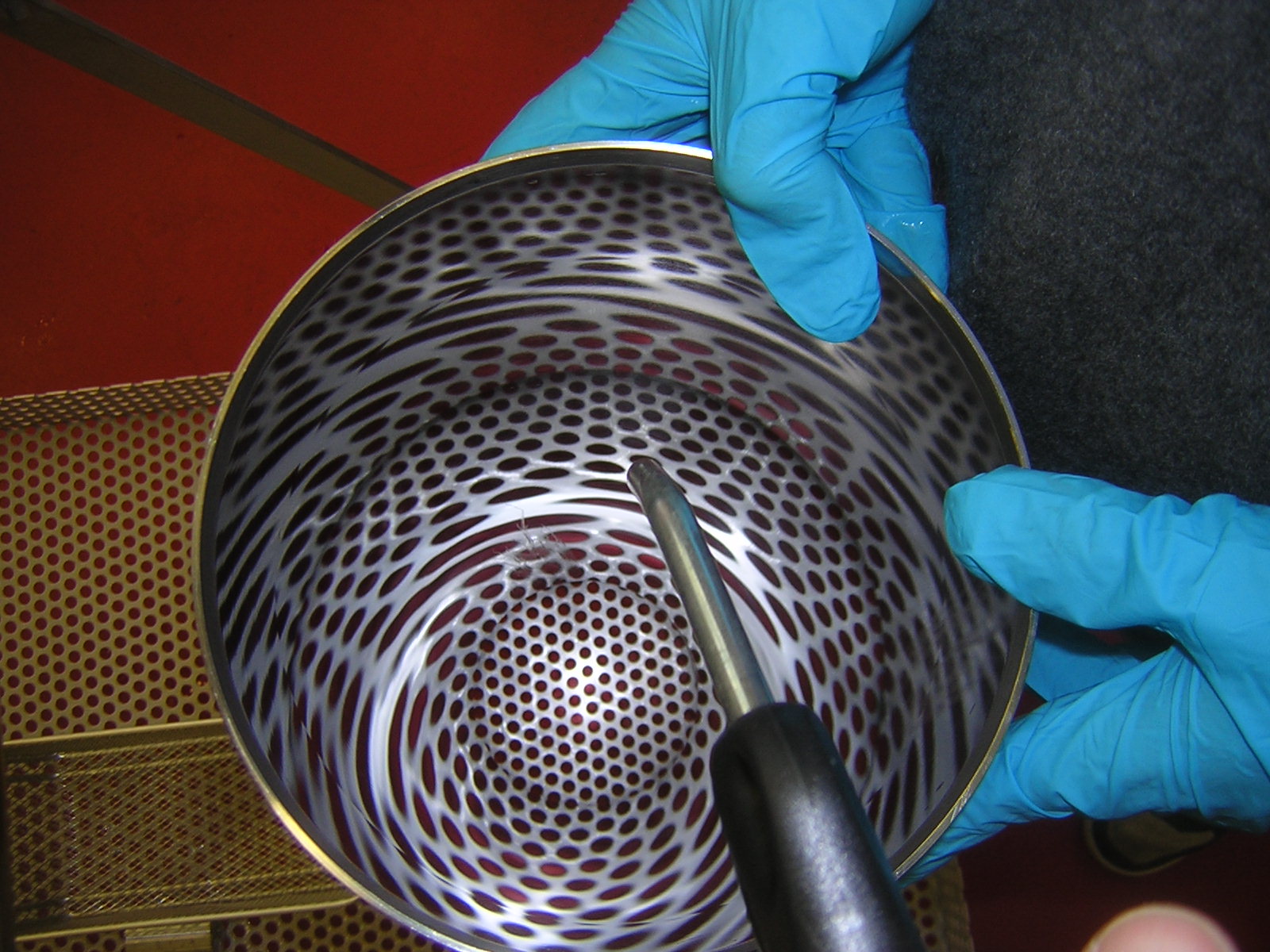}
}
\caption [Metal guide section] {
Photo of a stainless-steel guide during cleaning.
The optical reflection is an indication of the 
high surface quality.
}
\label{fig:metal-guide}
\end{center}
\end{figure}

\begin{figure}[htb]
\begin{center}
\resizebox{0.4\textwidth}{!}{\includegraphics{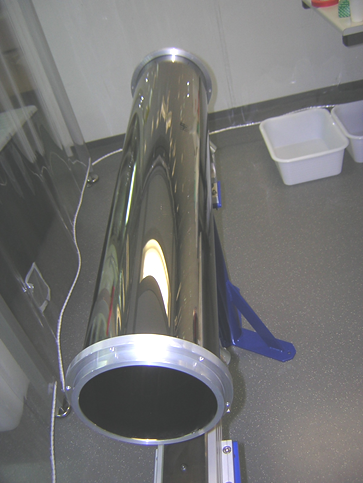}
}
\caption [Glass guide] {
Assembled UCN glass guide.
}
\label{fig:glass-guide}
\end{center}
\end{figure}

\begin{figure}[htb]
\begin{center}
\resizebox{0.4\textwidth}{!}{\includegraphics{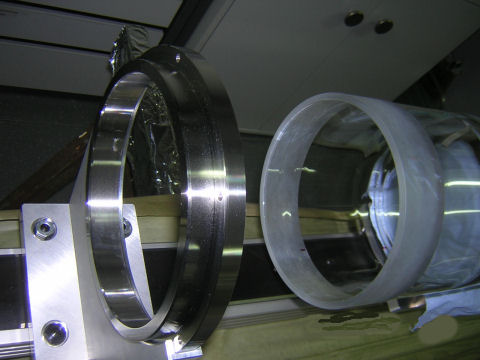}
}
\caption [Glass guide end] {
UCN glass guide end section before 
glueing the metal end-cap. 
The end of the glass tube was 
carefully ground and then flame-polished
to fit well the metal end-cap
in order to minimize gaps.
The reflections show the high surface finish.
}
\label{fig:glass-guide-end}
\end{center}
\end{figure}

\begin{figure}[htb]
\begin{center}
\resizebox{0.35\textwidth}{!}{\includegraphics{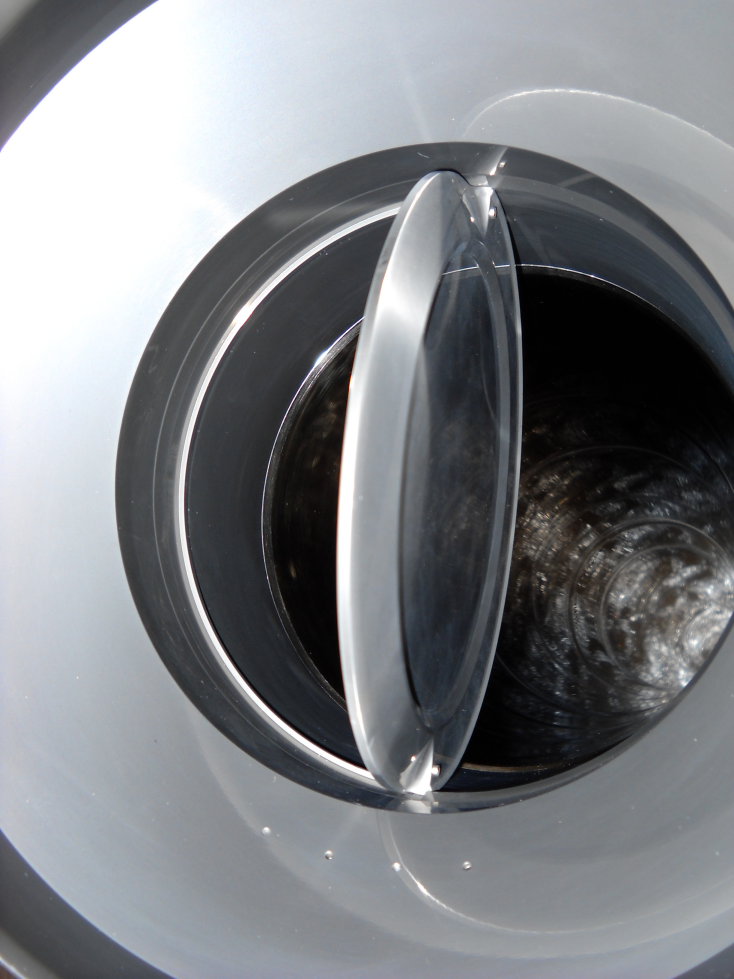}
}
\caption [Neutron guide shutter] {
Photo of the DLC-coated neutron guide shutter South
in open position from the inside of the storage vessel.
The visible four small openings on the bottom of the
tube served to position the  
endoscopic UCN counters of Ref.~\cite{Goeltl2013}.
}
\label{fig:nlk-south}
\end{center}
\end{figure}

All guide parts were either glued together,
(the radiation hardness of the glue was tested, see~\cite{Bertsch2009}),
or screwed tightly together,
in order to minimize gaps, 
which could cause significant UCN losses.
Figure~\ref{fig:metal-guide} shows a photo of one
metal guide.
Figure~\ref{fig:glass-guide} shows a picture of a finished glass guide.
Figure~\ref{fig:glass-guide-end} shows a photo of the end section
of a glass guide during the production process.
The end section was ground and smoothened over the entire diameter
to guarantee the dimensions and then flame-polished
to prevent chipping.
A stainless steel end-ring was glued onto the 
prepared glass guide 
providing good protection of the glass and minimizing gaps.
Any left-over small gap was filled with glue~\cite{Bertsch2009}
and then sputter-coated with the entire guide in one process.
The stainless steel guide attached to the storage vessel
serves also as thermal insulator 
between the storage vessel at 80\,K and 
the glass guide at room temperature.
It contains the neutron guide shutter at its front end 
shown in Fig.~\ref{fig:nlk-south}.

\begin{figure}[htb]
\begin{center}
\resizebox{0.8\textwidth}{!}{\includegraphics{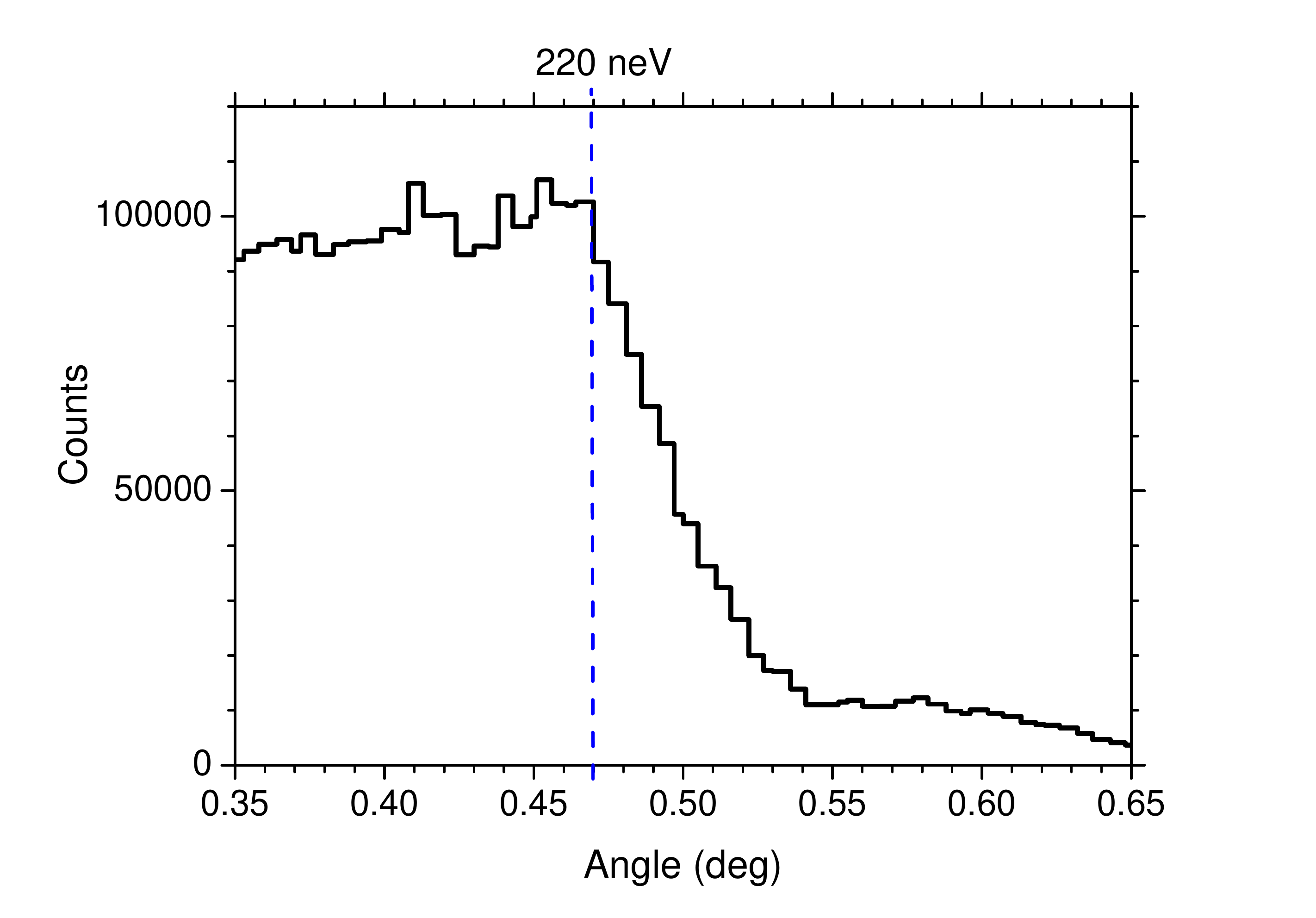}
}
\caption [NiMo reflectometry] {
Plot of a cold neutron reflectometry measurement
of one NiMo coating sample
showing reflected neutron counts versus reflection angle.
The vertical line indicates the angle 
corresponding to a neutron optical potential
of 220\,neV.
}
\label{fig:reflectometry}
\end{center}
\end{figure}

All surfaces exposed to UCN during operation
were finally sputter-coated with nickel-molybdenum (NiMo) with
a weight ratio of 85\% to 15\%.
Additional small flat glass pieces were attached to the tubes during the coating process
to allow cold-neutron reflectometry afterwards.
Measurements of all guide samples at the NARZISS reflectometer~\cite{Bodek2008}
resulted in a material optical potential of 220$\pm$10\,neV. 
One result of such a measurement of a NiMo coating 
is shown in Fig.~\ref{fig:reflectometry}.

The guide shutters and tubes indicated as No.
(6) in Fig.~\ref{fig:UCN-source},
and shown in Fig.~\ref{fig:nlk-south} for guide South,
are part of the storage vessel 
and therefore coated with DLC.
%
Opening and closing times of the guide shutters were measured to be below 200\,ms.

The UCN transmission of all installed guides was measured prior to installation 
and found to be about 98\% per meter
for all glass and stainless-steel tubes~\cite{Blau2016}.

\subsubsection{Minimizing gaps:}

\begin{figure}[htb]
\begin{center}
\resizebox{0.5\textwidth}{!}{\includegraphics{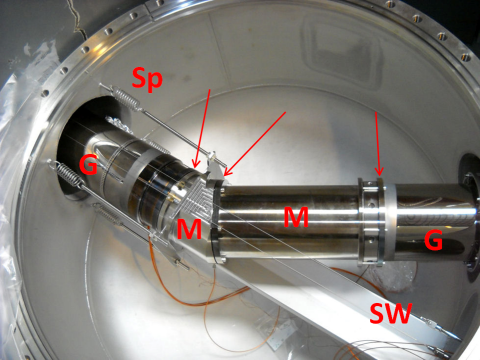}
}
\caption [Bent guide section] {
Photo of the bend section 
showing the connections (indicated with an arrow)
between metal guides (M) 
and glass guides (G).
The springs (Sp) pushing the front guide
towards the storage vessel are visible.
The two stainless steel wires (SW) on top of the guide are used to 
mechanically open and close the neutron guide shutter.
The optical cable serves as readout for a monitoring
detector~\cite{Goeltl2013}.
}
\label{fig:metal-guide-bend}
\end{center}
\end{figure}

Gaps along the UCN transport path can cause
large losses,
therefore we tried to minimize gaps
in the design and construction of the entire UCN path.
All metal flanges were precisely machined and glued onto the glass guides
as shown in Fig.~\ref{fig:glass-guide-end}.
The metal guides and glass guides were then firmly connected
by screws.
The front section, where the guide connects to the storage vessel
was pushed inside a concave receptacle on the storage vessel plate.
Strong mechanical springs were mounted to 
keep the tension to this guide section towards the
storage vessel
in order to stay firmly connected when the storage vessel is cooled 
to about 80\,K.
Otherwise, gaps with the size of the order of millimeters
could open due to thermal shrinking.
The bend section of guide West-1,
Fig.~\ref{fig:metal-guide-bend}, 
shows the
connections between metal guides and glass guides, 
and
also the four springs pushing the guides towards the storage vessel.
%
%
After installation the gap between guide end and window grid 
at the beamport was
measured to be 
0.3\,mm (West-1), 5\,mm (South), 0.1\,mm (West-2)
at room-temperature.

\subsection{Safety window}
\label{sec:safety-window}

\begin{figure}[htb]
\begin{center}
\begin{subfigure}[]
{\includegraphics[width=0.50\textwidth]{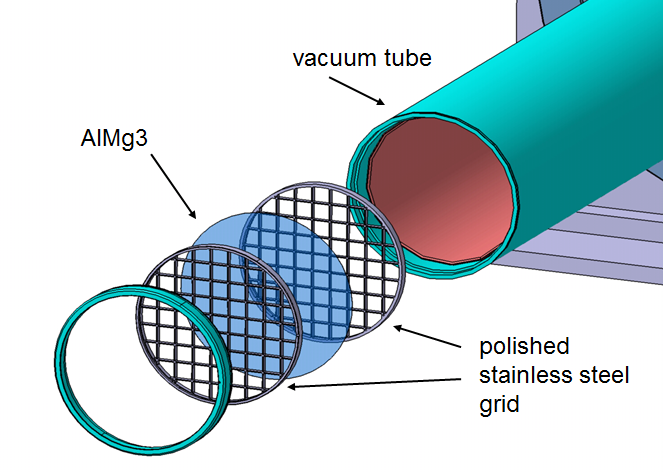}}
\end{subfigure}
\begin{subfigure}[]
{\includegraphics[width=0.40\textwidth]{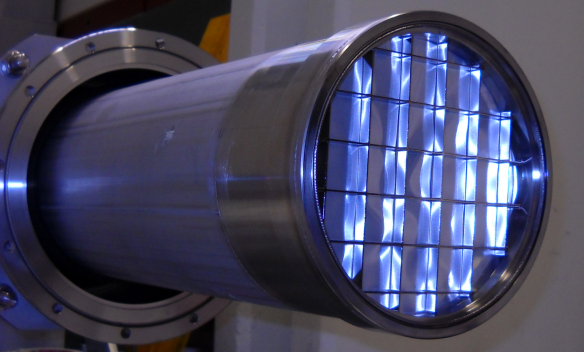}}
\end{subfigure}
%
%
\caption [Safety window] {
a) Expanded CAD image of the safety window section which separates UCN source vacuum 
from experiment vacuum. 
Shown are the two reinforcement grids, the AlMg3 foil,
the vacuum tube and holding ring. 
All the shown parts were finally electron-beam welded for tightness.
b) Photo of the installed safety window 
at the end of the vacuum tube of guide West-1.
}
\label{fig:safety-window}
\end{center}
\end{figure}

The safety windows at the end of the UCN guides
separate the UCN source vacuum from the experiment vacua.
Figure~\ref{fig:safety-window} 
shows all parts of the window assembly which were
finally welded together to assure vacuum tightness.
It is an important safety component and has therefore to withstand
3\,bar overpressure from inside and 1\,bar from outside.
After a long performance test series ~\cite{Atchison2009}, 
AlMg3 with a thickness of 0.1\,mm 
was selected as window material
because of its mechanical strength and 
good UCN transmission.
Measurements of UCN transmission in the AlMg3 foil
are reported in Section~\ref{sec:foil-transmission} below.

\begin{figure}[htb]
\begin{center}
\begin{subfigure}[]
{\includegraphics[width=0.45\textwidth]{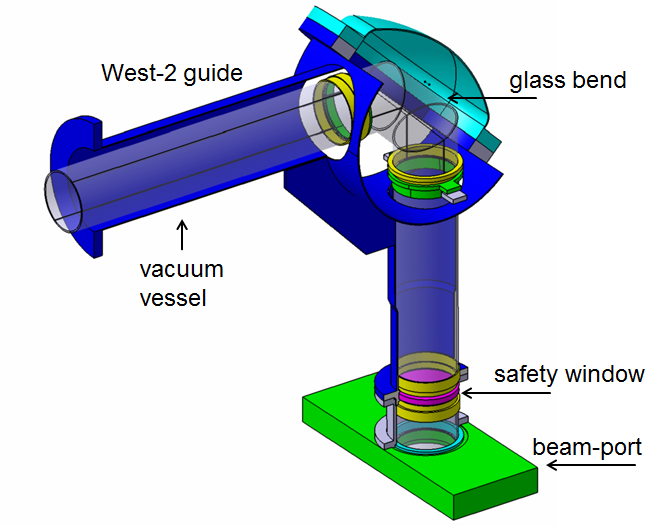}}
\end{subfigure}
\begin{subfigure}[]
{\includegraphics[width=0.40\textwidth]{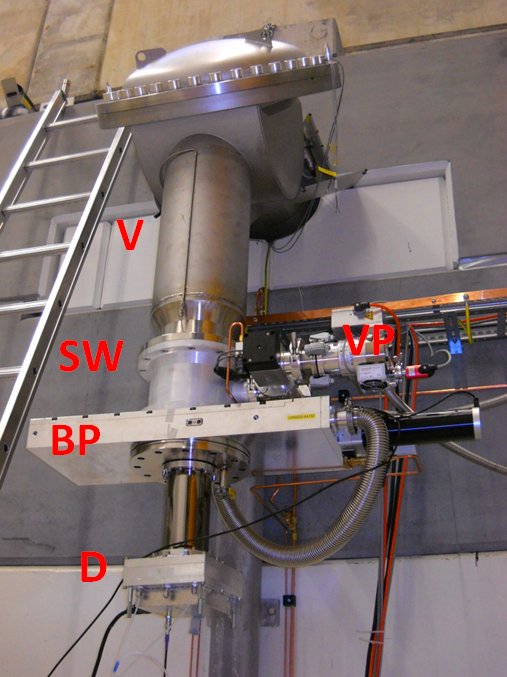}}
\end{subfigure}
\caption [West-2 bend section] {
a) CAD image of the vertical bend section of the West-2 guide
b) West-2 guide outside the biological shield
with positions indicated for:
vertical section (V),
safety window (SW),
beamport shutter (BP),
small Cascade detector (D),
vacuum pumps (VP).
}
\label{fig:West-2}
\end{center}
\end{figure}

All three guides accommodate a safety window located close to the beamport.
On guide West-2
the expected average UCN energy spectrum is significantly softer
due to gravitational shift, as
the extraction is at a height of 2.3\,m above the bottom of the storage vessel.
Therefore an additional vertical section on guide West-2,
shown in Fig.~\ref{fig:West-2}, was added.
This causes the UCN to fall by one meter and
increases the kinetic energy by 100\,neV,
well above the neutron optical potential
of the AlMg3 safety window (54\,neV).

The beamports West-1 and South were prepared for the installation of 
superconducting (SC) polarizer magnets. 
The safety windows would be located at the highest magnetic field of an installed SC magnet. 
The polarized UCN would have the highest velocities when passing the foil. 
For the nEDM experiment, beamport South was equipped 
with a magnet providing a 5\,T field. 
UCN intensity measurements for various magnetic field strengths 
are described in Section~\ref{sec:magnet} below.

\subsection{Beamport shutter}

Three large (DN200) vacuum shutters custom designed and produced 
by VAT\footnote{VAT Group AG, Seelistr. 1, CH-9469 Haag, Switzerland}
are mounted as beamport shutters indicated
as BP in Fig.~\ref{fig:West-2}.
All parts of the shutter exposed to UCN
were coated with DLC. 
In Ref.~\cite{Blau2016}  
we have already reported on the properties 
of this shutter.
Figures~4 to 6 of this reference depicted
the measured opening area and
opening function relevant for UCN transport modeling.
The opening and closing times were measured to be about 1\,s. 

\clearpage

\section{Simulation model for the PSI UCN optics}    
\label{sec:simulation}

In order to relate the parameters and the geometry of individual parts of the UCN source and guides system to its overall performance, detailed Monte Carlo simulations were performed. 
A large number of alternative geometry configurations for the UCN optics system
were simulated showing that the as-built configuration is optimal in terms of UCN yield.
The general features of the PSI-developed simulation code, MCUCN, for 
ultracold neutron transport are summarized in Ref.~\cite{Zsigmond2018}.
This code was also used during the development of the UCN source at PSI to optimize 
the distance between the sD$_2$ vessel and the bottom of the storage vessel. Also the size and shape of the storage vessel, and the guide diameters were investigated, to provide preliminary predictions on UCN density and energy distribution.

\subsection{Geometry model of the PSI UCN source}

In the geometrical model for MC simulations of the PSI UCN source we implemented (i) all surfaces coated with materials with high optical potentials (DLC, Ni, NiMo 85/15), and (ii) all gaps between guide sections which were approximated either as totally absorbing or having low material optical potential, e.g. aluminum.
As basis for the input geometry we used technical drawings.
The actual length of the neutron guides from storage vessel to beamport is implemented with an accuracy of 2 mm. 
The geometry starts with the lid of the solid deuterium vessel as shown 
in Fig.~\ref{fig:D2-vessel}.
However, instead of a half toroidal shape we implemented a dome. 
The latter definition was more straightforward 
and less run-time demanding
than building a toroidal surface 
from a series of cone sections.  Our previous MC 
comparisons of a dome and a toroidal shape gave the same results for the transmission of UCN because the average angular distributions of the surface normal are in both cases similar. 

This lid constitutes the lower boundary of the vertical guide 
which directs the neutrons into the large storage vessel of the UCN 
source (see Figs.~\ref{fig:storage-vessel} 
and \ref{fig:UCN-source}). 
This vessel has four flat sides. In the bottom part the two neutron guide exits are located in a niche.
Important are also the time dependent position of the flaps 
separating the storage vessel from 
the vertical guide, 
as shown in Fig.~\ref{fig:flapper-valve}.
The neutron guide shutter 
at the storage vessel is
shown for guide South in Fig.~\ref{fig:nlk-south}.
Apart from the lid of the sD$_2$ vessel, 
we also included the vacuum separation foils, 
made from AlMg3, at the end of the UCN guides as introduced 
in Sec.~\ref{sec:safety-window}.
On guide South the foil was placed 
in the center of a 5\,Tesla superconducting magnet 
which acts as UCN polarizer. 
Each virtual detector counts UCN through another, similar 
100\,$\mu$m thick separation foil.
The sensitive detector area is matched to the detector used 
in the actual measurement.
For test calculations different experimental volumes were attached
depending on the aim of the calculation.

\begin{figure}[htb]
\begin{center}
\resizebox{0.49\textwidth}{!}{\includegraphics{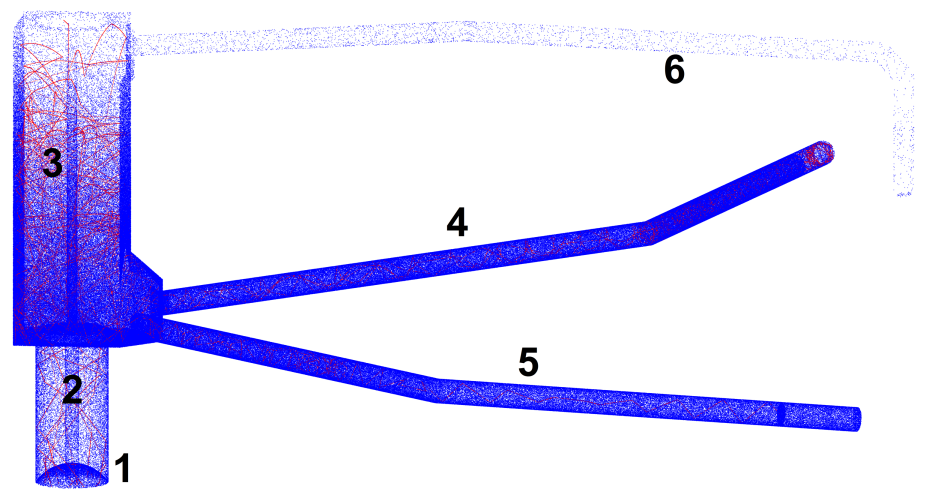}
}
\caption [Simulation model] {
Simulation model of the UCN source visualized by example trajectories (red) and reflection points of 
UCN (blue) on the relevant neutron optics surfaces: 1 - lid separating the UCN source vessel from the vertical neutron guide, 2 - vertical neutron guide, 3 - UCN storage vessel, 4 - UCN guide South, 5 - UCN guide West-1, 6 - UCN guide West-2. The density of the reflection points is proportional to the number of wall interactions. Thus such 3D geometry visualizations serve at the same time as first visual tests of the physics implementation.
}
\label{fig:Simulation-model}
\end{center}
\end{figure}

\begin{figure}[htb]
\begin{center}
\resizebox{0.49\textwidth}{!}{\includegraphics{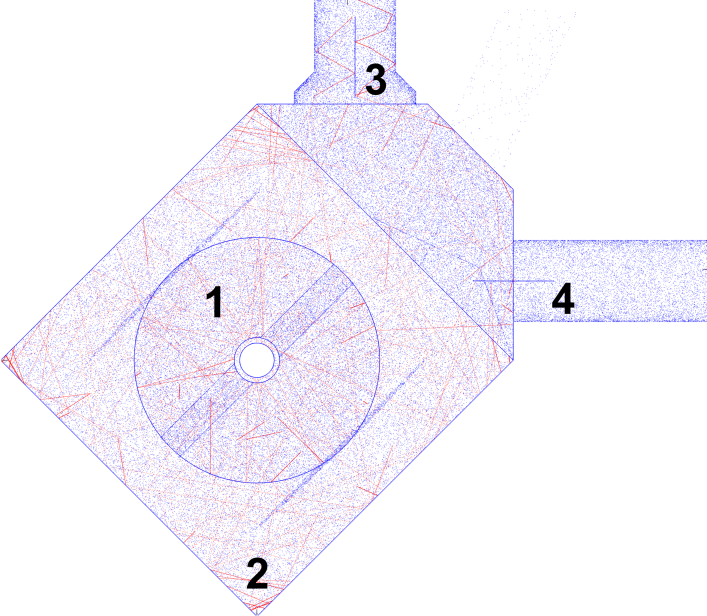}
}
\caption [Simulation model - zoom] {
View from the top of the simulated geometry visualized by example trajectories (red) and reflection points of 
UCN (blue) depicting details of the storage vessel: 1 - top of the vertical guide (large circle) , 2 - storage vessel, 3 - neutron guide ``South'', 4 - neutron guide ``West-2''. The reflection points on the main flapper valves are plotted for both the closed state (within large circle) and the opened state (two narrow `clouds' tangent to the large circle). 
}
\label{fig:Simulation-model-zoom}
\end{center}
\end{figure}

In Figs.~\ref{fig:Simulation-model} and~\ref{fig:Simulation-model-zoom} 
the geometry of the
UCN source part as appearing in the MCUCN model
is visualized.  The blue dots are points of reflection of UCN on their way to the beamports and example trajectories are shown in red.
Such plots serve as first tests of the geometry and physics implementations.

\subsection{Coating parameters}

We introduced three global coating parameters for the neutron guides between the storage vessel and the beamports:
\begin{enumerate}
	\item an optical potential with a value obtained 
from cold neutron reflectometry measurements~\cite{Atchison2006,Atchison2006b,Atchison2007a}; 
	\item a loss parameter $\eta$, calculated from the ratio of the complex and real parts of the optical potential - independent of kinetic energy in agreement with experimental results~\cite{Atchison2007a,Bondar2017}; 
	\item a parameter p$_{diff}$ for the fraction of diffuse reflections according to the Lambert model 
which is also independent of kinetic energy. 
\end{enumerate}
The history of the trajectories in the MC simulations show that the majority of UCN exiting the beamports have sampled the whole volume of the UCN source and the beamline and therefore we use such global i.e. spatially averaged parameters (the spacial averaging is of course not independent from the geometry configuration).
Similarly, different values of the three global parameters were defined for the entire storage vessel of the UCN source, 
including shutters, and separately for the vertical guide between UCN converter and storage vessel of the source.

\subsection{Initial trajectory parameters from the source}
\label{sec:InitialTraj}

At the beginning of the simulation UCNs are generated with random initial coordinates (position, velocity and time) as described below according to probability distributions detailed in~\cite{Golub1991}.

For the energy distribution of the UCN emerging 
from the solid deuterium surface, 
we rely on a linear approximation, i.e.
on the initial linear section of a Maxwell flux (eq.(3.9) in~\cite{Golub1991})
for a high-quality, realistic sD$_2$ converter. 
This assumption was strengthened by separate MC simulations 
of the UCN extraction efficiency from bulk sD$_2$
for a close-to-optimal sD$_2$ structure. 
They are in agreement with experimental data~\cite{Altarev2008}.
These calculations included (i) super-thermal UCN production, 
(ii) temperature-dependent thermal up-scattering, 
(iii) up-scattering on para-deuterium, 
(iv) Porod scattering~\cite{Brys2007}, 
(v) absorption on hydrogen contamination, and 
(vi) reflection or absorption on the coating material of the converter vessel. 
We did not implement an interface distribution of macroscopic cracks 
in the solid deuterium since these are unknown and can be different 
after every freezing procedure.
Such cracks represent deuterium-vacuum interfaces which can reflect UCN 
and hence increase their dwell time within the material and thus also the loss probability.
We also did not implement intensity reduction due to the observed frost 
formation~\cite{Anghel2018}.
Consequently, our estimate of the absolute UCN flux constituted an upper limit 
and is not subject of this paper.

The angular distribution of the emerging UCN was set as a linear dependence 
in cosine of the angle to the surface normal. 
This is typical for moderators and for perfectly diffuse emitter models. 
On top of this angular distribution we added a vertical boost 
which corresponds to the optical potential of 
the sD$_2$~\cite{Daum2008}. 
Since we don't expect a perfectly smooth deuterium surface for a real converter 
material we also modeled cases when this boost was not vertical but diffuse. 
In this case we kept the initial angular distribution mentioned above and only increased the total energy by 102.5\,neV. Then the vertical transmission through the lid decreased by about 30\%.

The initial position coordinates are generated according 
to a flat horizontal homogeneous distribution confirmed 
by our separate MC results on extraction from the sD2 into vacuum. 
Deviations from this distribution don't show a relevant influence. 
%


\section{Characterization measurements}
\label{sec:characterization}


A series of measurements were performed to characterize the UCN source
components.
Measurement were performed with
UCN counters attached 
to beamport West-1 and West-2.
Beamport South was used for a few measurements only as
it permanently hosted
the nEDM experiment~\cite{Pendlebury2015,Baker2011, ABel2018a}
up to 2017 and 
has the n2EDM experiment~\cite{Abel2018b} now being set up.

We used position-sensitive Cascade UCN 
counters\footnote{
CDT CASCADE Detector Technologies GmbH,
Hans-Bunte-Str. 8-10, D-69123 Heidelberg, Germany}
with active areas of 10\,cm$\times$10\,cm (referred to later as `small') or
20\,cm$\times$20\,cm (`big'), the latter covering the full 180\,mm diameter
of the beam guide.
UCN detection works via neutron capture
on $^{10}$B which is followed by emission of an $\alpha$ particle and 
a $^7$Li nucleus~\cite{Klein2011}.
The detector entrance window,
a 0.1\,mm thick AlMg3 or Al foil
is coated on the backside 
with a $^{10}$B layer of 200\,nm.
The UCN transmission in the Al part of this entrance window 
is about 0.7~\cite{Goeltl2008}.

In standard operation, UCN counts are monitored with the big cascade 
detector mounted on beamport West-1 
and one small cascade detector mounted on beamport West-2.

\subsection{UCN counts time distributions}
\label{sec:emptying}

The time distribution of UCN counts arriving
at a given beamport reflects the 
emptying time of
the storage vessel and UCN guide
through the defined opening 
of the neutron guide
towards a detector.
The time behavior can be described with an exponential function
and a characteristic time constant ($\tau$) which for storage measurements is 
called storage time constant (STC)
\begin{equation}
N(t) = I \times e^{\frac{-t}{\tau}}     
\end{equation}
\label{eq:expdecay}

\begin{figure}[htb]
\begin{center}
\resizebox{0.70\textwidth}{!}{\includegraphics{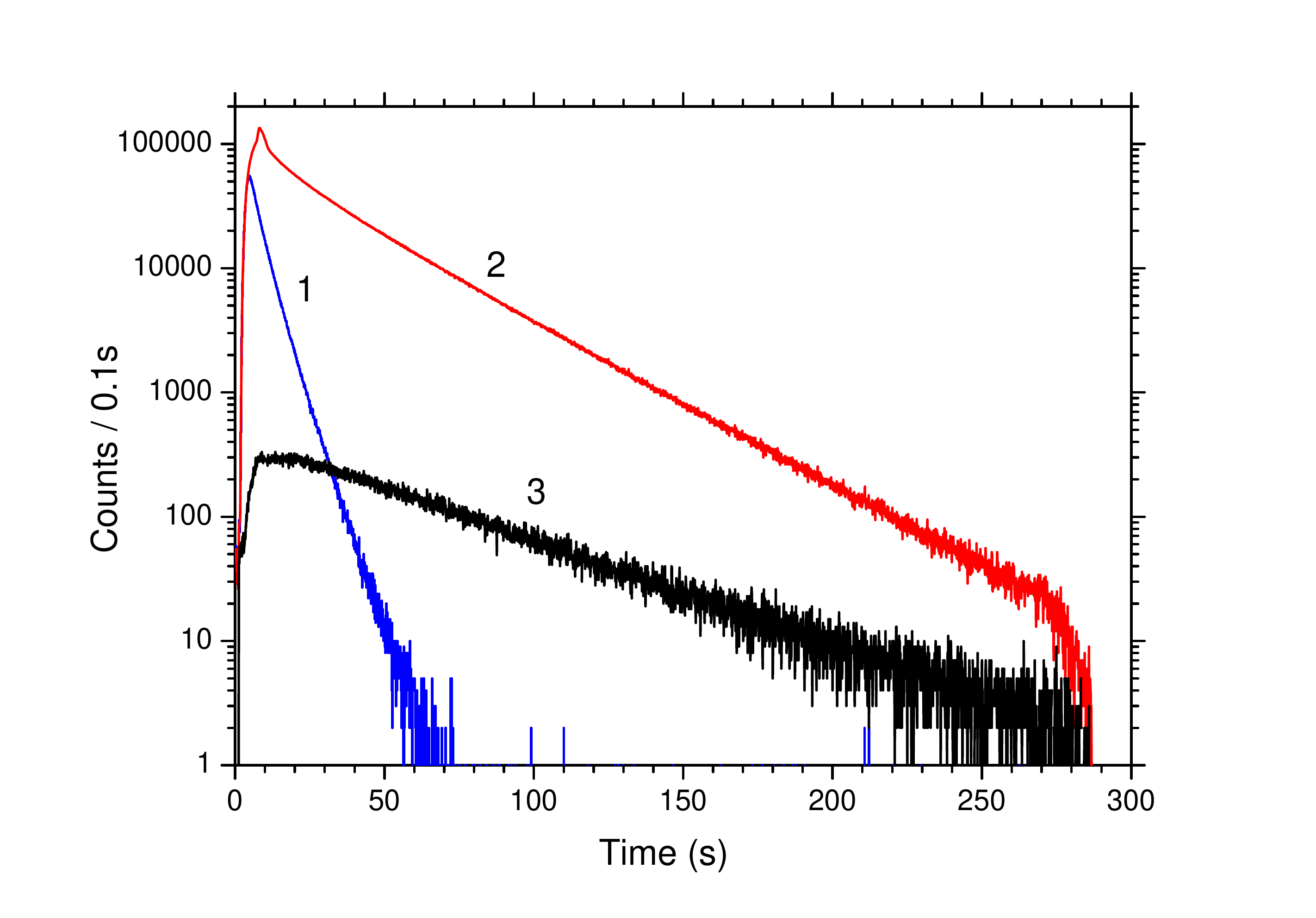}
}
\caption [UCN counts West-1] {
UCN counts observed at beamport West-1 for three different
operation modes:
1) benchmark pulse ($\tau$ = 5.0\,s);
2) production pulse ($\tau$ = 36.0\,s);
3) leakage operation with neutron guide shutter closed ($\tau$ = 90\,s).
The bin width is 100\,ms, values from Tab.\ref{table:emptying-times}.
}
\label{fig:UCN-West-1}
\end{center}
\end{figure}

We mainly worked with three configurations:
Figure~\ref{fig:UCN-West-1} shows the observed UCN counts 
over 
a full period of 300\,s between proton pulses 
for three different operation modes,
i.e.\ configurations of the UCN source.\newline
1) Benchmark pulse: A benchmark pulse is a 2\,s long proton beam pulse 
with a box-shaped profile. All shutters are open and hence no
mechanical movements influence UCN delivery. This mode is
used in order to guarantee 
close to identical conditions for measurements distributed over many years.
As the storage vessel flapper valve stays open, UCN are rapidly lost
due to back-reflections through the large opening 
(a factor of 9 larger area than the guide exits) 
and subsequent absorption in the sD$_2$ solid.\newline
2) Production pulse: 
In standard operation, i.e. providing production pulses, the
UCN storage flapper valves close
at the end of the proton pulse
preventing further back-reflection towards the sD$_2$.
UCN are then emptied towards the beamport over the
full period between subsequent pulses. 
The UCN count rate rapidly increases during the pulse.
After the flapper valve is closed the rate decreases with 
the emptying rate of the storage vessel and guide system
through one or in this example two
guide openings, 254\,cm$^2$ each.
At 280\,s the flapper valve opens again in preparation of the next
beam pulse, therefore the remaining UCN are quickly lost. \newline
3) Leakage mode: 
In leakage mode the neutron guide shutter towards
the beamport stays closed. 
The shutter is not perfectly tight 
and UCN can leak through and travel 
from the exit of the storage vessel
to the detector.
The initial emptying rate is smaller
and has a longer time constant compared to case (2).
This mode is used to determine the 
storage time constant of the storage vessel.
At 280\,s the flapper valve opens.

\begin{figure}[htb]
\begin{center}
\resizebox{0.70\textwidth}{!}{\includegraphics{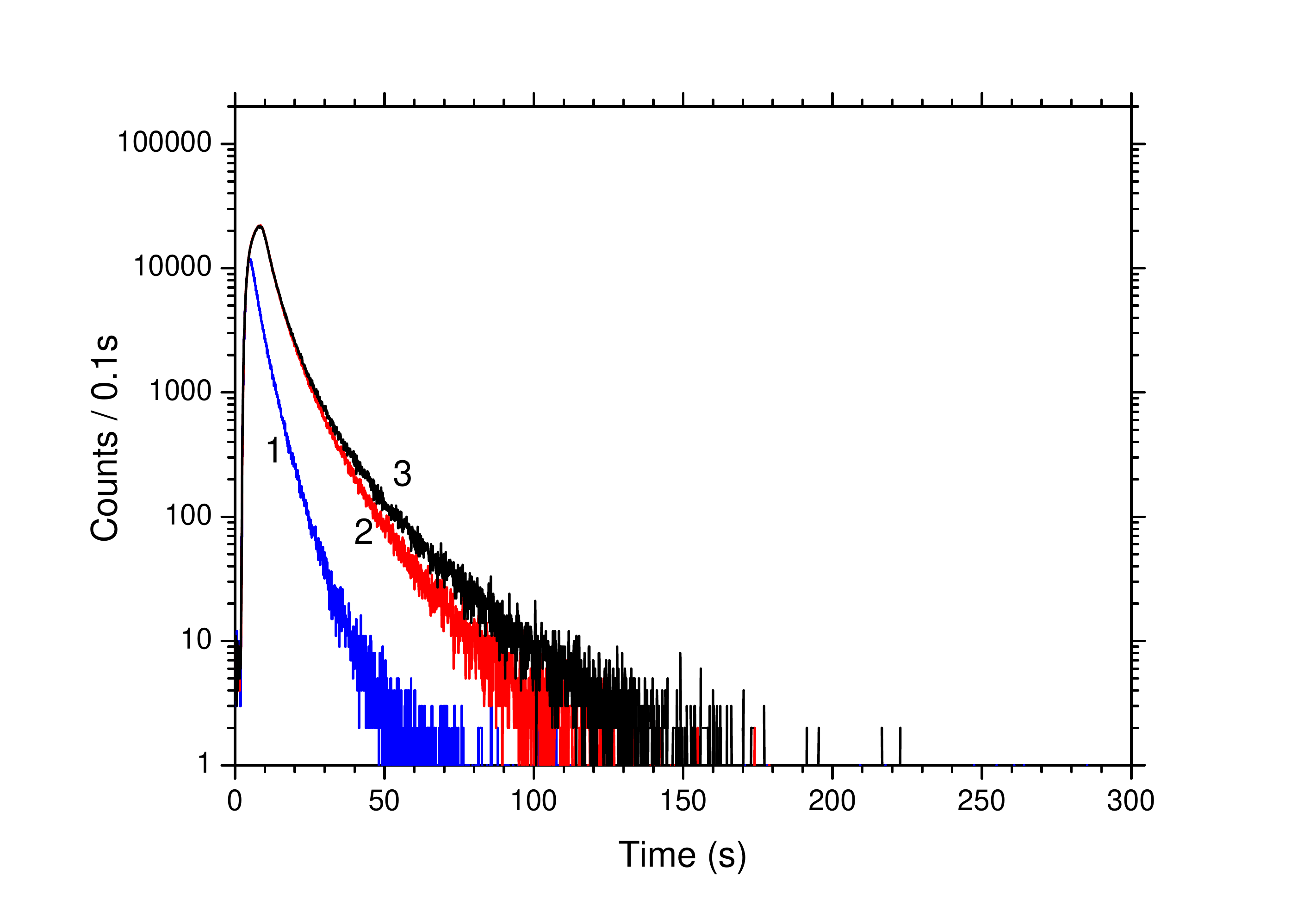}
}
\caption [UCN counts West-2] {
Same as Fig.~\ref{fig:UCN-West-1}
but observed at beamport West-2.
Bin width is 100\,ms.
}
\label{fig:UCN-West-2}
\end{center}
\end{figure}

Figure~\ref{fig:UCN-West-2} displays the 
same three cases as in Fig.~\ref{fig:UCN-West-1},
but at beamport West-2.
The UCN intensity is about a factor of 10 smaller
due to 
i) the extraction height of guide West-2
which is 2.3\,m above the storage vessel floor
and
ii) using the small Cascade counter.
After about 100\,s all UCN with energies above 230\,neV 
at the level of the West-2 storage-vessel exit
are extracted or lost,
and the UCN intensity is rapidly approaching zero.
Also the observed storage time constants are
smaller than at beamport West-1 as the observed UCN
which can reach the West-2 beamport
have on average higher energies
at the bottom level
and therefore are lost faster. 
The fact that neutron guide shutter West-1 is closed,
influences the UCN intensity at West-2
only at very late times
as can be seen by the higher intensity of curve 3) 
in comparison to curve 2).

Depending on the 
open or closed state of the main flapper valves of
the storage vessel and the different neutron guide shutters
the storage time constants differ.
The resulting characteristic time constants 
from fits of Eqn.\ref{eq:expdecay} to the data
are listed in Tab.~\ref{table:emptying-times}.

\begin{table}
\centering
\begin{tabular}{|c|c|}
  \hline
  Source Configuration & $\tau$ [s] \\ \hline
\multicolumn{2}{|c|}{SV-Flaps closed}\\
  \hline
West-1 open        & 36.1$\pm$0.2 \\
W-1 \& South open  & 31.6$\pm$0.7 \\
  \hline
\multicolumn{2}{|c|}{SV-Flaps open}\\
  \hline
  2 flaps open    & 5.0$\pm$0.4 \\
  one flap closed & 6.9$\pm$0.2 \\
  \hline
\end{tabular}
\caption[Emptying times]{
List of storage time constants measured with a detector on beamport West-1 
and with different shutter configurations (see also~\cite{Goeltl2012}).
The shutter to West-2 was always open and does not significantly 
influence the time constants.
}
\label{table:emptying-times}
\end{table}

\subsection{UCN intensity development}
\label{sec:UCN-intensity}

\begin{figure}[htb]
\begin{center}
\resizebox{0.70\textwidth}{!}{\includegraphics{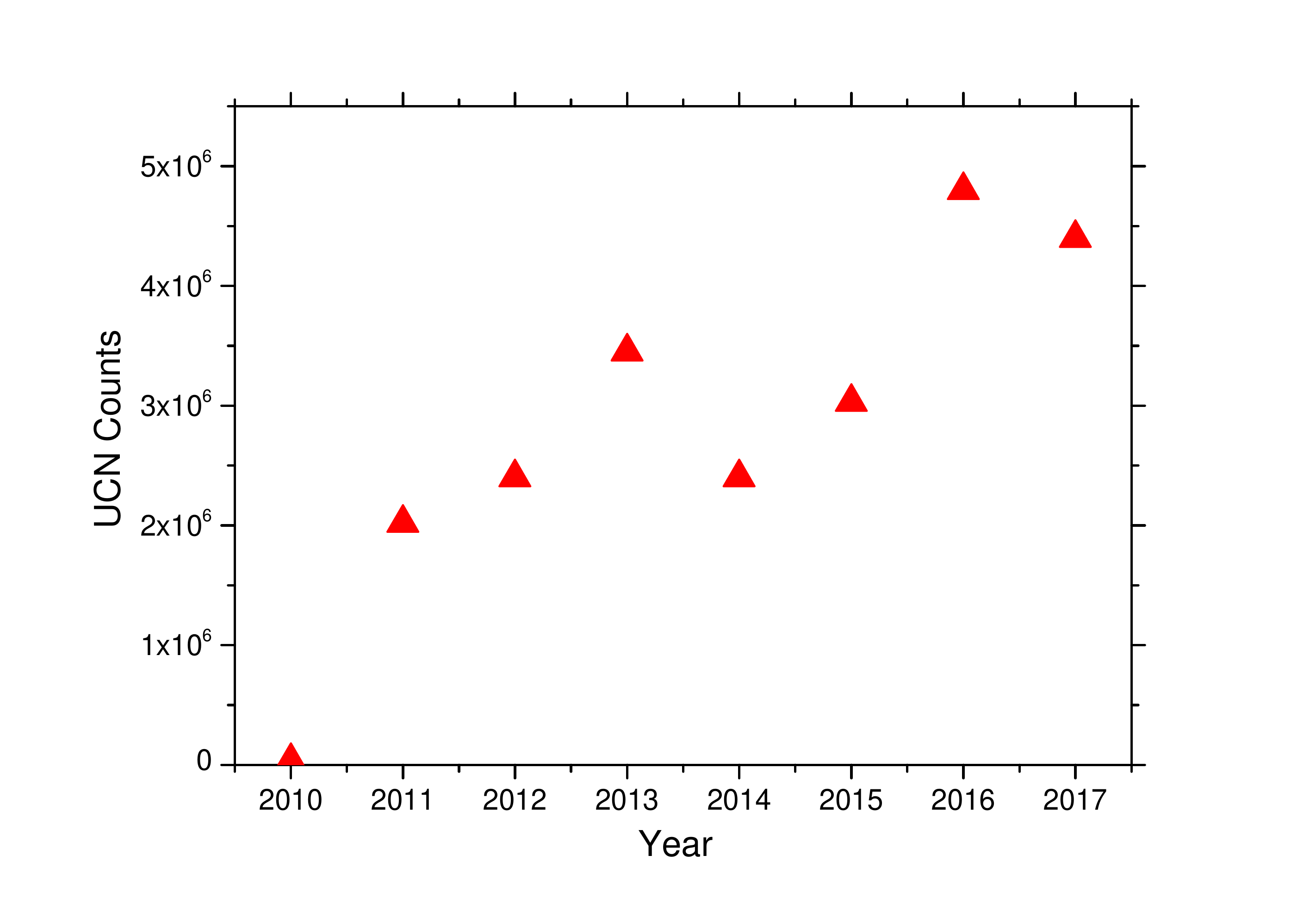}
}
\caption [Benchmark pulse history] {
Highest UCN output for a benchmark pulse observed on beamport West-1
in a given year, normalized to the proton beam 
current of 2.2\,mA.
}
\label{fig:max-benchmark-pulse}
\end{center}
\end{figure}

\begin{figure}[htb]
\begin{center}
\resizebox{0.70\textwidth}{!}{\includegraphics{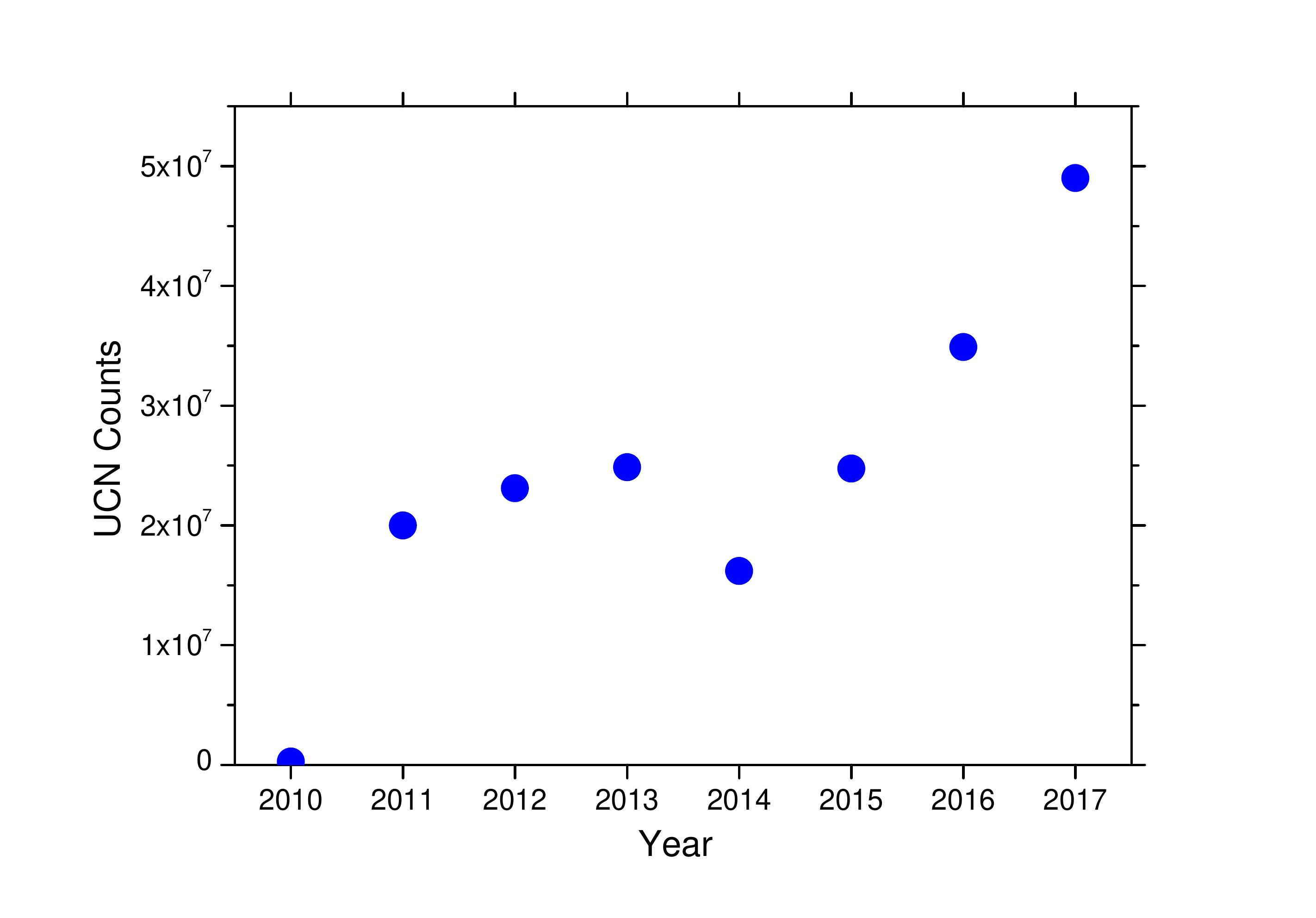}
}
\caption [Production pulse history] {
Highest UCN output for a production pulse observed on beamport West-1
in a given year, 
normalized to the proton beam
current of 2.2\,mA.
}
\label{fig:max-production-pulse}
\end{center}
\end{figure}

In order to characterize the intensity of the UCN source
and compare it over long time periods 
we have used benchmark pulses and production pulses.
Benchmark pulses do not involve any mechanical movements 
and should therefore track the status of the UCN production and
extraction in the sD$_2$, 
under the assumption of no changes in 
UCN transport through storage vessel, guides and windows.
The observed intensity increase over the years, shown 
in Fig.~\ref{fig:max-benchmark-pulse},
tracks the improvements
in preparation and modification of the sD$_2$
and the increased duty cycle.
In Fig.~\ref{fig:max-production-pulse} 
the largest UCN yield measured from
production pulses on beamport West-1 is
shown for each year.
%
Actual measurements took place at proton beam currents 
of 1.8, 2.2, 2.4\,mA 
and were normalized to 2.2mA for the plot.
%
Both plots show a dip in 2014 because a 
different experimental setup was online~\cite{Bondar2017}
when the UCN source actually performed best, 
hence no standardized characterization measurement was
taken at the best source conditions.

\subsection{Storage time constants of the UCN storage vessel}
\label{sec:storage}

The storage time constant of the storage vessel can be determined
in two ways. 
The 'classical` method~\cite{Golub1991} 
proceeds in three steps:

\begin{enumerate}

\item filling the storage vessel with UCN; 

\item closing all UCN shutters and storing the UCN for a given time;

\item opening the shutter towards the detector and counting all surviving UCN.

\end{enumerate}
The leakage method as described in Sec.~\ref{sec:leakage-method}
is an alternative way to measure the decrease of the UCN density
in the storage vessel.
With both methods the time dependence of the observed UCN intensity
can be fitted with a single (or double, or multiple) 
exponential function
with the slope defining the storage time constant(s) (STC).

\subsubsection{Classical method}

\begin{figure}[htb]
\begin{center}
\resizebox{0.70\textwidth}{!}{\includegraphics{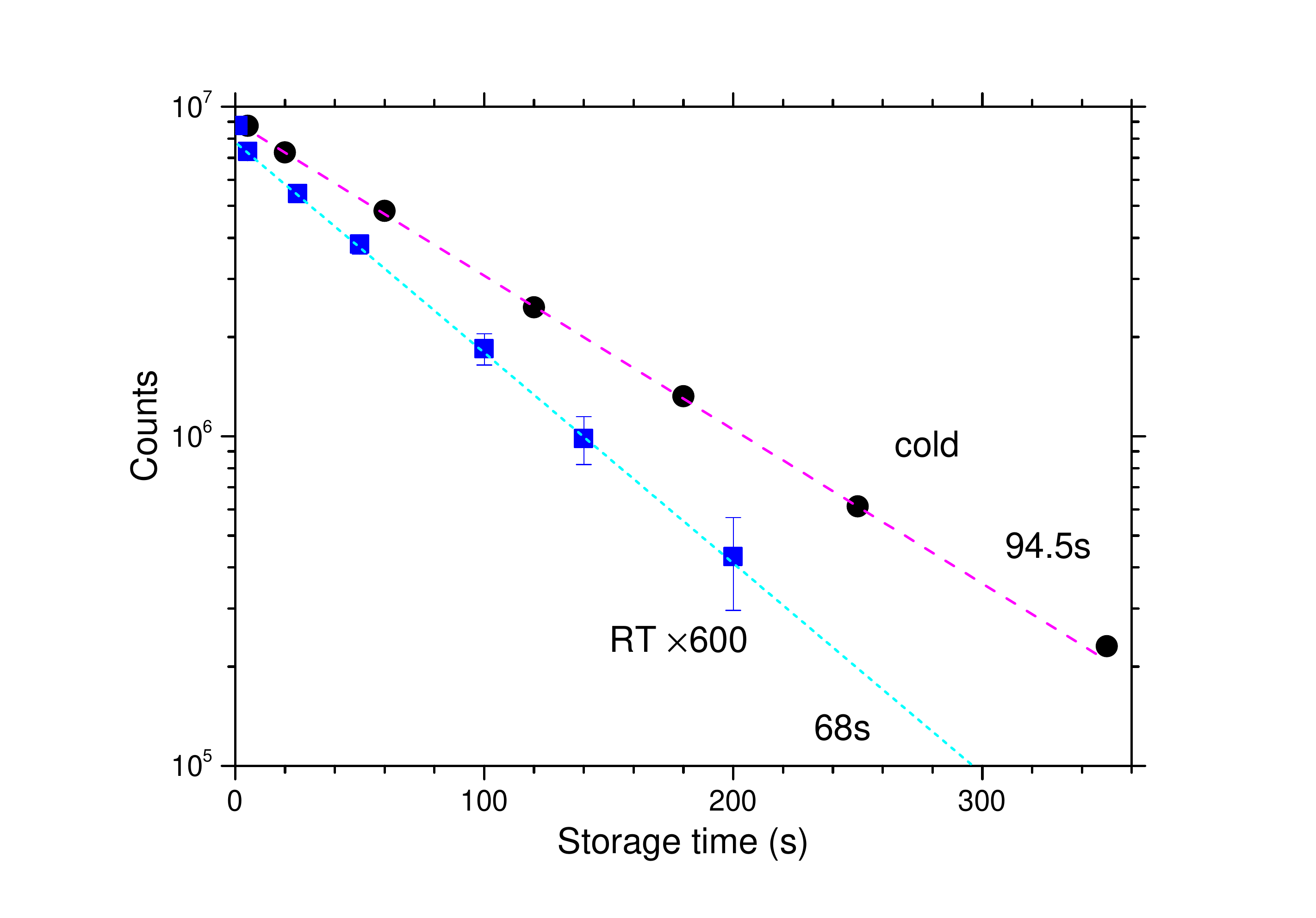}
}
\caption [Storage time cold] {
UCN counts observed after a given storage time in the storage vessel:
a) at operating conditions around 80\,K in 2011.
The dashed line guides the eye with a single exponential decay 
function and a storage time constant of 94.5\,s. 
b)
at room temperature with gaseous D$_2$ for UCN production.
The dotted line shows a fir with STC=68\,s.
}
\label{fig:storage-classic-cold-RT}
\end{center}
\end{figure}

Measurements using the classical method 
were performed with a Cascade counter mounted on 
beamport West-1.
We have measured the number of UCN in the vessel as a function of
time for storage times between 5 and 350~s. 
By choosing different
fit regions one can obtain information about the behavior 
of the storage vessel with respect to the UCN velocity spectrum.
A fit to the entire data-set yields a STC~=~94.5$\pm$0.9~s for a 
single exponential as shown in Fig.\ref{fig:storage-classic-cold-RT}.
In principle every UCN energy group has a separate storage time constant.
It is common to define two groups which are discernible 
with separate storage time constants.
The short time range 
is dominated by the influence of spectral cleaning 
which takes some time until
UCN with kinetic energies above the optical potential of the walls
are lost.
But it also characterizes the group of faster UCN.
Since these interact more often with the wall materials
they are lost faster.
After longer times only slower UCN survive resulting in a larger STC.
Table~\ref{table:STC} lists the result obtained for different fit ranges.

\begin{table}
\centering
\begin{tabular}{|c|c|}
  \hline
  Fit range (s) & STC (s) \\
	\hline
5 - 300 &  94.5$\pm$0.9  \\ 
\hline
5 - 20 &  80.0$\pm$10.0  \\
180 - 350 &  97.2$\pm$0.5  \\
  \hline
\end{tabular}
\caption[Storage time constant]{
Storage time constants from single exponential fits to the storage 
measurement at cold operating conditions in 2011.
(see also~\cite{Goeltl2012}).
}
\label{table:STC}
\end{table}

In order to understand the temperature dependence
of losses due to thermal up-scattering, 
we also performed a measurement with the 
storage vessel at room temperature.
This could only be done with very low UCN statistics
and in a non-standard operation mode where the
entire UCN source was at room temperature.
The UCN intensity was therefore very low, 
as the UCN production took place in 
gaseous D$_2$ at room-temperature and 1\,bar pressure, 
hence under very disfavoring conditions.
The results are plotted in Fig.~\ref{fig:storage-classic-cold-RT}.
A STC of 68\,s was determined, with the at 2\,s measurement 
outside the fit range.



\subsubsection{Leakage method}
\label{sec:leakage-method}

The `leakage method' makes use of the fact that
the neutron guide shutters are not perfectly UCN tight 
but cause a small UCN leakage on the order of 1\%.
The intensity of these leaking UCN
decreases exponentially
with time described with the STC
of the UCN storage vessel.
%
%
The measurement is shown in 
Fig.~\ref{fig:UCN-West-1} by curve 3
which was fitted 
with a single exponential function.
The results confirm the classical measurement.

Several measurements were conducted in subsequent years of operation
which were separated by periods of 
warming up the entire facility to room temperature, 
and venting of the storage vessel and guides.
Furthermore, some maintenance work was typically performed 
during such shutdown periods.
The measured STCs
with a single exponential fit
in an identical fit range,
given in Tab.~\ref{table:STC-per-year},
agree within a small range of a few seconds.
The lower values observed in 2013 could not
be conclusively explained.
The small changes could be due to variations in 
closing of the flapper valve of the storage vessel,
in closing of the neutron guide shutters,
or to the UCN energy spectrum from changing
conditions of the sD$_2$
bulk and surface.
A change in vacuum conditions, i.e. rest gas, is excluded.

\begin{table}
\centering
\begin{tabular}{|c|c|}
  \hline
  Year & STC (s) \\\hline
2011  &  89$\pm$1  \\ \hline
2012  &  84$\pm$1  \\ \hline
2013  &  79$\pm$1  \\ \hline
2014  &  83$\pm$1  \\ \hline
2017  &  83$\pm$1  \\ \hline
2018  &  85$\pm$1  \\
\hline
\end{tabular}
\caption[Storage time constant]{
Storage time constants from exponential fit to the storage 
measurement at cold operating conditions in different years
since start-up.
Data were binned to 1\,s and a
fit in the range of 40 to 220\,s after proton pulse start~\cite{Ries2016}.
}
\label{table:STC-per-year}
\end{table}

\subsection{Transmission through AlMg3}
\label{sec:foil-transmission}

UCN have to penetrate solid AlMg3 material 
on their path from production to detection
at two positions 
which are required by hydrogen safety regulations:\newline
1) the lid of the solid deuterium moderator vessel (thickness = 500$\mu$m)
(see Fig.~\ref{fig:D2-vessel});\newline
2) the UCN vacuum window at the end of the UCN guides (thickness = 100$\mu$m)
(see Fig.~\ref{fig:safety-window}).\newline

It was previously shown that measured transmission through Al foils cannot 
be correctly calculated 
by relying only on the Al absorption 
cross-sections and material densities which 
consistently overestimates the transmission~\cite{Atchison2009}.
Therefore we have pursued several investigations into the UCN transmission 
properties of the AlMg3 material used in the foils.

Measurements were performed using a Cascade detector mounted 
after a foil holder in the UCN path.
Different foils
with 70\,mm diameter were inserted.
The setup was installed on beamports West-1 and West-2.
The two positions allowed to see the influence of 
different UCN energy spectra on the transmission.

1) Production pulse on West-2: 
The West-2 port extracts UCN from the top of the storage vessel
at 230\,cm height.
After a 1.3\,m vertical section before the beamport 
(see Fig.~\ref{fig:West-2})
all UCN have a minimum
energy of $\sim$130\,neV due to gravitational acceleration.
Two 90-degree UCN guide bends made from stainless steel 
after the vertical section were used to cut off the UCN
with kinetic energies above the stainless steel 
neutron optical potential $\sim$190\,neV.

2) Production pulse on West-1: 
At this port UCN energies start at 50\,neV 
(passing the safety foil) 
and are well populated
up to the
guide coating 
material optical potential of 220\,neV.\newline
A fraction of higher energy neutrons is also present 
as shown in MCUCN simulations 
(see Fig.~\ref{fig:EnergySpectrumEvolutionWest-1DirectTransmission}).
In the following we discuss the two types of safety windows.

\begin{figure}[htb]
\begin{center}
\resizebox{0.70\textwidth}{!}{\includegraphics{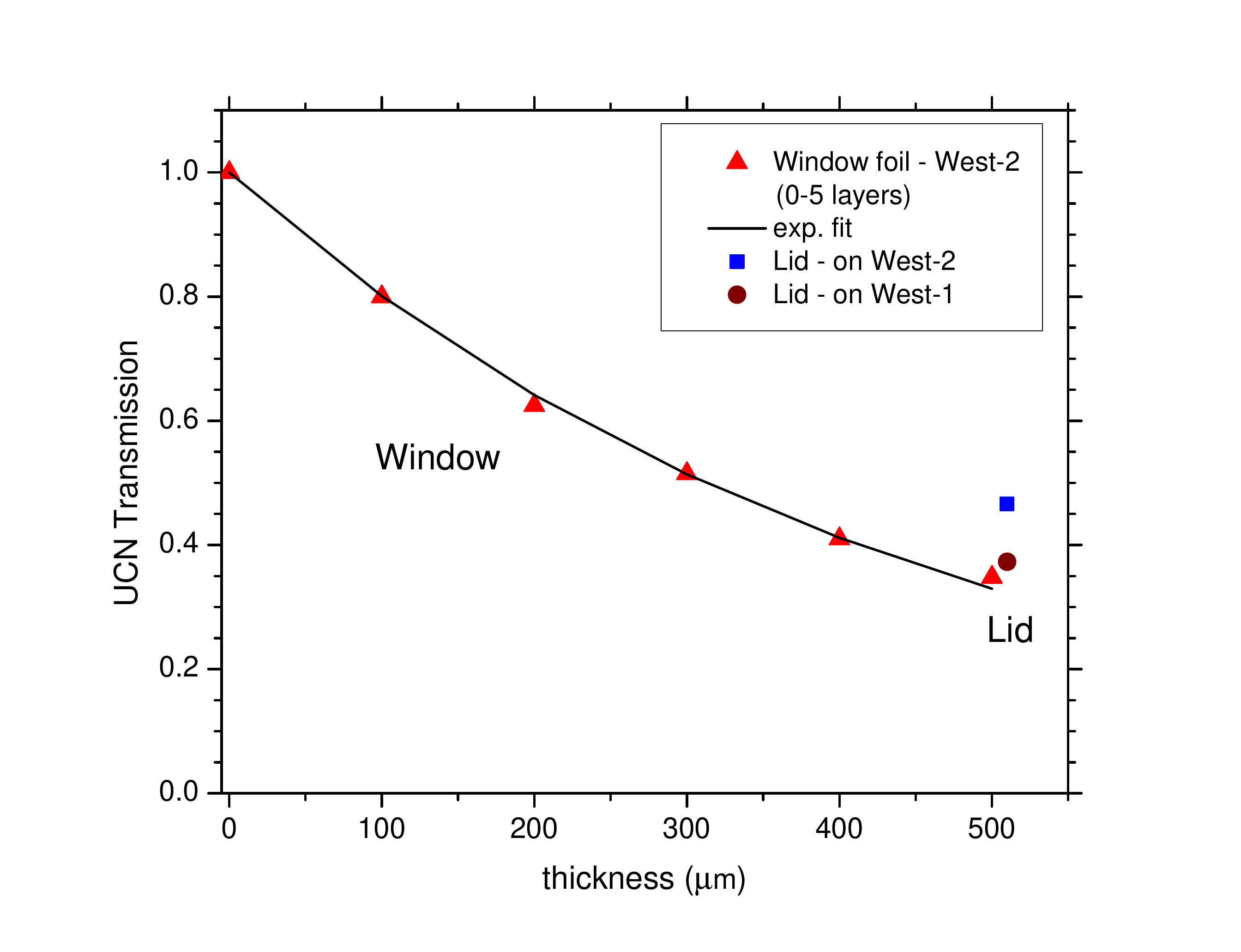}
}
\caption [AlMg3 transmission]  {
Measured UCN transmission for up to 5 layers of the
window safety foil (filled triangles) and 
the moderator vessel lid (filled circle and square) 
at various beam conditions. 
The line represents a fit to the foil measurement 
taking into account the shape of the simulated energy spectrum in West-1. 
This fit yielded a factor of 2.1$\pm$0.1 multiplying the theoretical 
loss cross-section (see text Sec.~\ref{sec:simulation-analysis}).
}
\label{fig:foil-transmission}
\end{center}
\end{figure}

\subsubsection{D$_2$ moderator vessel lid}
\label{sec:lid}

The 0.5$\pm$0.05\,mm thick material used in the lid
measurement was machined in the same way and 
from the same bulk AlMg3 material
as the installed moderator vessel lid.
This was possible
as the moderator vessel had an identical
twin which passed a burst-pressure test
demonstrating functionality up to 7\,bar.
After this burst test, large undamaged parts of 
the damaged lid could be re-used for UCN transmission measurements
at the West-2 and West-1 beamports.
The results are plotted in Fig.~\ref{fig:foil-transmission}.
%
It is obvious, that a lower UCN kinetic energy reduces the
UCN transmission. 
We remind the reader
that UCN gain about 100\,neV in kinetic energy
when exiting from the sD$_2$ due to its neutron optical 
potential~\cite{Daum2008}.
Therefore, the UCN energies at the lid position starts at 100\,neV.
The height difference of the West-1 and West-2 beamports
selects a different UCN energy spectrum.
%
%
The measured values are consistent with one previously obtained
using a similar Al-alloy~\cite{Goeltl2008}
measured at the TRIGA Mainz UCN facility~\cite{Kahlenberg2017}.

\subsubsection{Window foil}
\label{sec:window}

The material for the UCN transmission measurement
of the window safety foil 
was taken from the same 0.1\,mm thick AlMg3 foil roll 
as the original window foil used for the installed
safety windows (see Sec.~\ref{sec:safety-window}).

Measurements on West-2 were performed 
with  
a stack of up to five foils,
demonstrating 
the expected exponential decrease with thickness 
displayed in Fig.~\ref{fig:foil-transmission}.
One can see that while having the same material
thickness of 0.5\,mm, five foils together have a smaller UCN transmission 
than a single 'thick' piece.
This is probably
due to additional reflections on the individual foil surfaces.

Calculating the UCN transmission from absorption cross-section 
values for known material compositions results 
typically in a large underestimation of the UCN transmission
as already observed previously in~\cite{Atchison2009}.
Therefore experimental transmission values 
were used to tune the 
loss cross-sections in the Monte-Carlo simulation.
It required increasing the theoretical cross-section
by about a factor of factor of 2.1$\pm$0.1 
consistent with the findings in Ref.~\cite{Atchison2009}.
These effective cross-sections can then reproduce the
observed UCN transmission values.
These were used in our simulation 
model of the PSI UCN source as outlined in Sec.~\ref{sec:simulation}.
%


\subsection{Transmission measurements through the superconducting magnet}
\label{sec:magnet}

As described in Section 2.5, 
beamport South was equipped with a SC magnet. 
Because of the magnetic moment of the neutron interacting with the 
magnetic field ($\pm$60\,neV/T) 
the UCN velocity of the transported UCN is affected in a magnetic field 
and therefore also when passing this AlMg3 window. 
We have performed transmission measurements for various field 
values in order to further constrain the UCN energy spectrum from the source.  
The big Cascade detector was mounted
directly at that beamport
to investigate 
the window transmission as a function of magnetic field.
No additional spin analysis system was used. 
With this setup the dependence of the 
UCN rate for various magnetic fields 
in the range of \SI{0}{\tesla} to \SI{5}{\tesla},
using production pulses with \SI{3}{\second} length 
and benchmark pulses with \SI{2}{\second} were measured.
%


\subsubsection{UCN intensities}

\begin{figure}[htb]
\begin{center}
\resizebox{0.70\textwidth}{!}{\includegraphics{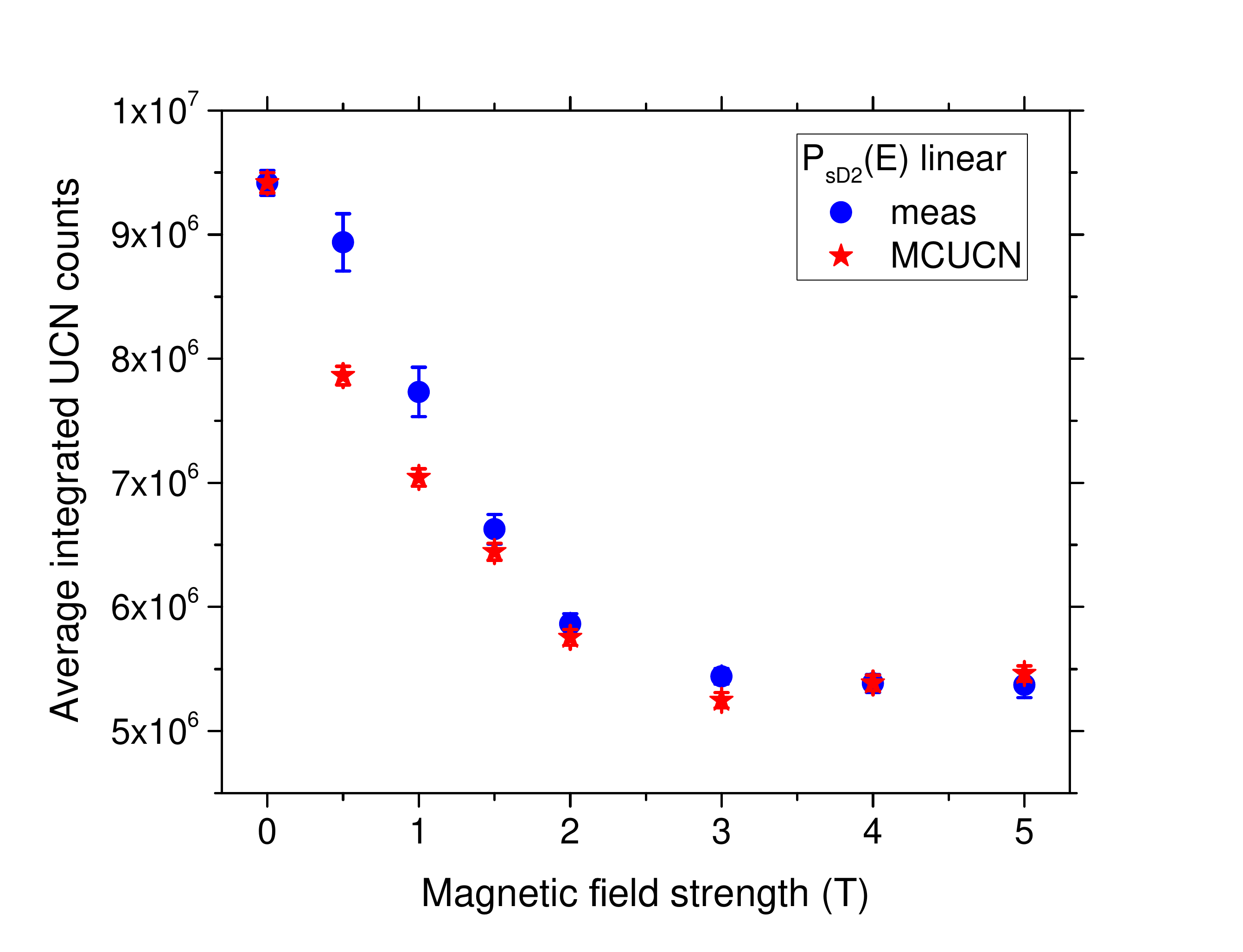}
}
\caption 
[Total UCN counts vs magnetic field] 
{
UCN counts for different SC magnet fields integrated 
from 1\,s
after the end of the proton beam kick 
up to 300\,s.
Measured values (filled circles) are
averaged over 3 proton beam kicks.
Filled stars mark the intensities resulting
from the simulation. %
 with a linear energy spectrum from the sD$_2$.
The difference at 0.5 and 1\,T shows
that the low energy range of the UCN spectrum is not yet well modeled as 
this is just above the Al threshold of the safety window.
}
\label{fig:totalUCNvsB}
\end{center}
\end{figure}

Integrated UCN counts for the different magnetic fields, 
measured in 3\,s production pulses are shown in 
Fig.~\ref{fig:totalUCNvsB}.
Table~\ref{tab:UCN-transmission-SCM} lists the intensities, both,
for benchmark pulses and production pulses.
In order to exclude faster neutrons occurring during the proton pulse,
the integration time started 1\,s after the end of the pulse.
With increasing magnetic field magnitude, lower rates of UCN were registered 
in the detector. 
This is expected, as half of the neutrons 
-- i.e. one spin state -- start
to encounter a potential barrier increasing with magnetic field 
strength (low field seekers),
while the other half (high field seekers) encounter a magnetic well, 
The lowest UCN intensities at $B=$\SI{5}{\tesla} 
are at 57\% of the ones at zero magnetic field. 
That this fraction is considerably larger than one half is 
because of the improved transmission of the transported spin state. 
These UCN have about 300\,neV more kinetic energy 
when passing the AlMg3 window than at B=0.


\begin{table}[htb]
\begin{center}
\begin{tabular}{|l|l|l|}
\hline
     B(T) & Transmission in 3\,s    & Transmission in  \\
					& production pulse        & benchmark pulse   \\ \hline
    0.0   &   \SI{1.00}{}        &  \SI{1.00}{}   \\
    0.5   &   \SI{0.95\pm0.03}{} &  \SI{0.958\pm0.003}{}  \\ \hline
    1.0   &   \SI{0.82\pm0.02}{} &  \SI{0.851\pm0.005}{}  \\ 
    1.5   &   \SI{0.70\pm0.01}{} &  \SI{0.744\pm0.006}{}  \\ \hline
    2.0   &   \SI{0.62\pm0.01}{} &  \SI{0.656\pm0.003}{}  \\ 
    3.0   &   \SI{0.58\pm0.01}{} &  \SI{0.580\pm0.002}{}  \\ \hline
    4.0   &   \SI{0.57\pm0.01}{} &  \SI{0.566\pm0.003}{}  \\
    5.0   &   \SI{0.57\pm0.01}{} &  \SI{0.565\pm0.002}{}  \\ \hline
\end{tabular}
\caption[UCN counts at different magnetic fields]{
UCN transmission with respect to magnetic field strength
relative to B=\SI{0}{\tesla},
measured with production pulses and benchmark pulses
a proton beam current of \SI{2200}{\micro\ampere}.
The larger error in the production pulse measurement
reflects the large time jitter of the flapper 
valve at the time of the measurements. 
This was largely improved at a later time~\cite{Ries2016}.
}
\label{tab:UCN-transmission-SCM}
\end{center}
\end{table}


\subsubsection{UCN rate measurements}

\begin{figure}[htb]
\begin{center}
\resizebox{0.70\textwidth}{!}{\includegraphics{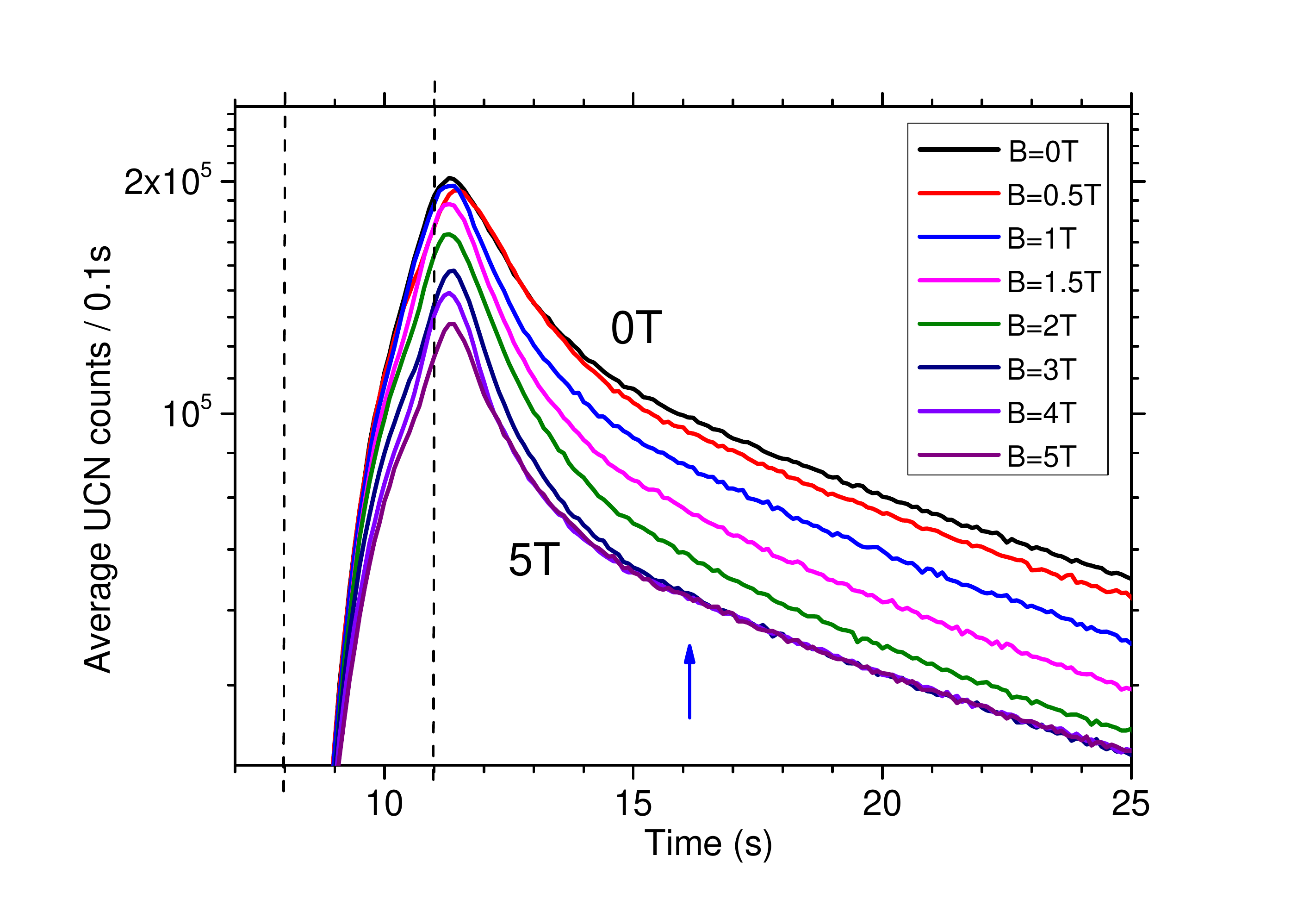}
}
\caption [UCN rate vs magnetic field] {
Display of
UCN rates per \SI{100}{\milli\second} bins for the first seconds after the proton pulse.
The vertical dashed lines indicate start and end times of the 
\SI{3}{\second} long proton beam kick.
The arrow indicates when the rates at
$B=$3, 4 and \SI{5}{\tesla} merge.
}
\label{fig:UCNvsBzoom}
\end{center}
\end{figure}

Looking into the first \SI{30}{\second} 
of the 
UCN intensity decrease after production pulses,
as shown 
in Fig.~\ref{fig:UCNvsBzoom}, 
reveals some information about the UCN energy spectrum.

\begin{enumerate}

\item The peaks of the count rates, about \SI{1}{\second} 
after the end of the proton pulse, are separated in intensity, as expected,
with intensities decreasing with increasing B field.
High field seekers 
and low field seekers
with forward momenta high enough to traverse the magnetic field
are exiting the beamport.
 
\item A few 
seconds after the end of the kick, 
the UCN rates for 
$B=$3, 4 and \SI{5}{\tesla} merge.
This means, that no high field seekers
with forward momenta 
high enough to traverse $B=\SI{4}{\tesla}$ magnetic fields
are left in the source.
A little bit later the same happens
for UCNs which cannot traverse $B=\SI{3}{\tesla}$.
At about \SI{50}{\second} after the end of the 
kick (not shown in Fig.~\ref{fig:UCNvsBzoom}), 
the same happens for UCNs which cannot traverse $B=\SI{2}{\tesla}$ .
\end{enumerate}

The measurements hence confirms the expected
higher loss rates of faster UCN.

\begin{figure}[htb]
\begin{center}
\resizebox{0.70\textwidth}{!}{\includegraphics{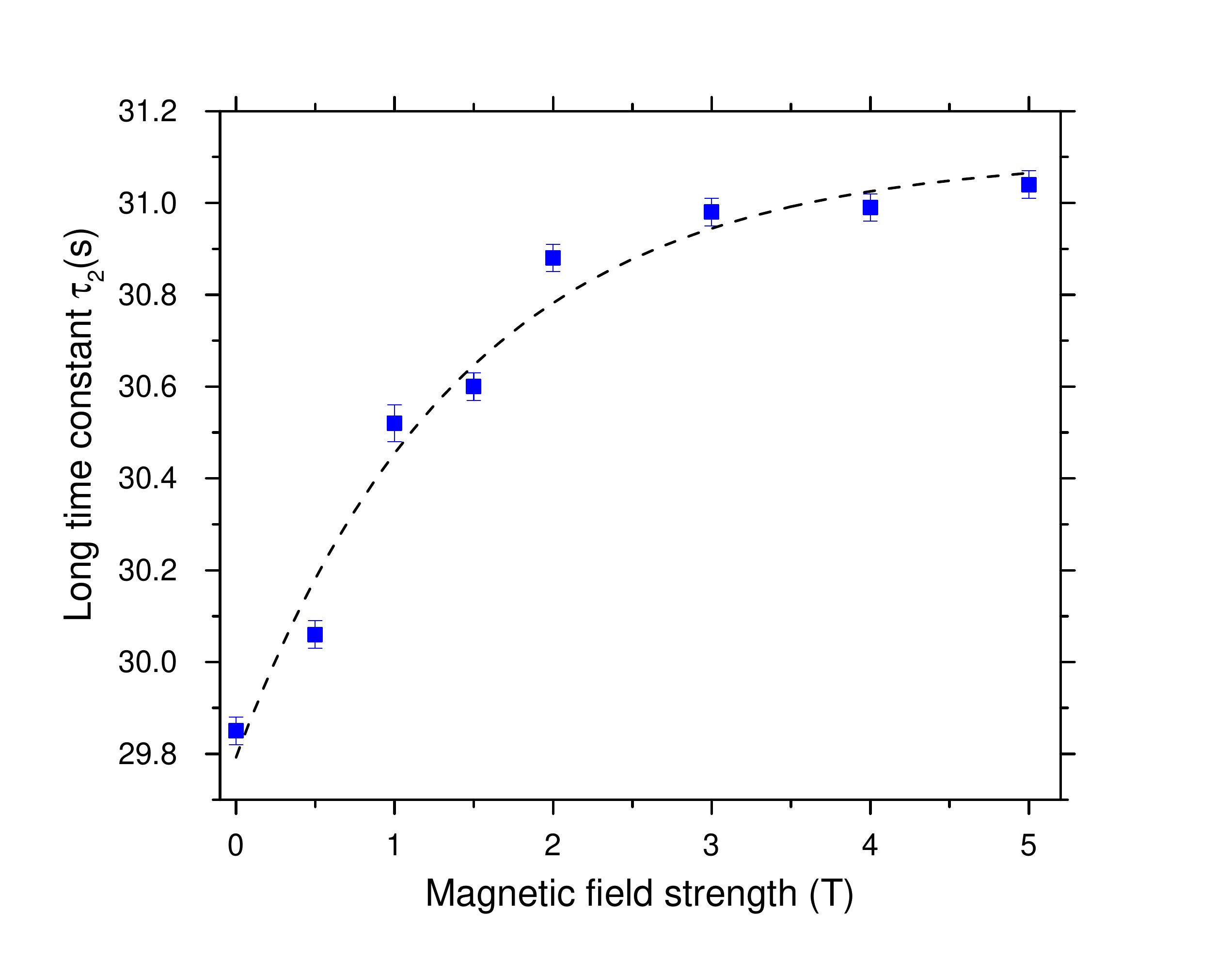}
}
\caption [Emptying time vs magnetic field] {
Long emptying time constant $\tau_2$ 
of the UCN source for different magnetic fields of the polarizing magnet.
The values and errors are extracted from fitting double exponential decay curves to the data.
The dashed line is to guide the eye.
}
\label{fig:EmptyingTimevsB}
\end{center}
\end{figure}

A double exponential decay of the form 
 
\begin{equation}
 N(t) = a e^{-t/\tau_1} + b e^{-t/\tau_2}  ,
\end{equation}

where $N(t)$ is the amount of UCN counted in a given time bin, $a$ and $b$
decay amplitudes for the short ($\tau_1$) and long ($\tau_2$) time constants,
was fitted to the time spectra.
The long time constant is shown in 
Fig.~\ref{fig:EmptyingTimevsB}
as a function of the magnetic field.
The fits were done in a time window from \SI{15}{\second} to \SI{300}{\second}, 
which corresponds to a \SI{285}{\second} long window starting at about
\SI{4}{\second} after the end of the proton beam kick.

The long time constant, of the order of \SI{30}{\second}, rises with
the magnetic field strength and saturates at \SI{3}{\tesla}.
This is likely because for slower UCN the
size of the potential barrier above 3\,T does not matter much.
In addition
the magnetic field reflects 
low field seekers
back into direction to the source.
There UCNs can
change the spin orientation, e.g. in a wall collision, and then return to the
magnet after a diffuse reflection from a wall or from the storage vessel.
Like this, some UCN exit the source at later times 
than in the absence of a magnetic field.
This leads to a longer emptying time.
A more detailed discussion is given in Sec.~\ref{sec:mag-simulation}.

\section{Simulation-based analysis of the measurements}
\label{sec:simulation-analysis}

The scope of the present MC analysis is to extract model parameters 
defined in the established theory of UCN physics 
in order to characterize the quality of the neutron-optical components, 
and to identify possible improvements. 
In all results presented in this paper, the simulated counts represent sampling trajectories  and we used free scaling parameters in order to best fit the experimental curves. Computed predictions of the absolute UCN counts at the PSI source will be the subject of a separate forthcoming study.

Our model on diffuse reflections, taking a fraction of reflections as ideally
diffuse (Lambertian), is only an approximation. 
However, for longer time scales which allow for a large number of reflections, storage-effects become important and 
the concrete mechanism of diffuse reflection will be less relevant. 
This was demonstrated in several test simulations using alternative 
diffuse reflection models. Such long time scales are for example our typical emptying and filling times in the experiments comparable to several tens of seconds.

In this sense, in our definition of diffuse reflection properties 
the diffuse fraction parameter $p_{\text{diff}}$ is an 
effective characterization quantity in terms of the Lambert model. 
We also performed simulations comparing the Lambert and 
micro-roughness~\cite{Steyerl1972b} models, 
e.g. for the case of emptying UCN from a storage chamber into a detector. We obtained almost identical results when setting the 
Lambert parameter accordingly low (to match micro-roughness) - as a consequence that storage effects were in this case dominant.


Different coating parameters influence with different weights 
the outcome of measurements or simulations. 
In some cases it was enough to scan one parameter and compare 
to the experimental data, the rest of the parameters being irrelevant. 
For example the calculated storage time constant of the source storage vessel, 
measured by the counts after UCN traversing the guides and exiting the beamport, 
is of the order of 100\,s, 
and dominated by the loss parameter.
%
We checked this by performing consistency calculations by varying the diffuse 
reflection parameters. 
In order to extract the UCN guide parameters multidimensional scans were 
necessary. In each case, single and multi-parameter scans, 
we also performed several 
consistency scans by roughly varying parameters 
which were expected to affect the result less.

\subsection{Storage vessel parameters}

\begin{figure}[htb]
\begin{center}
\resizebox{0.49\textwidth}{!}{\includegraphics{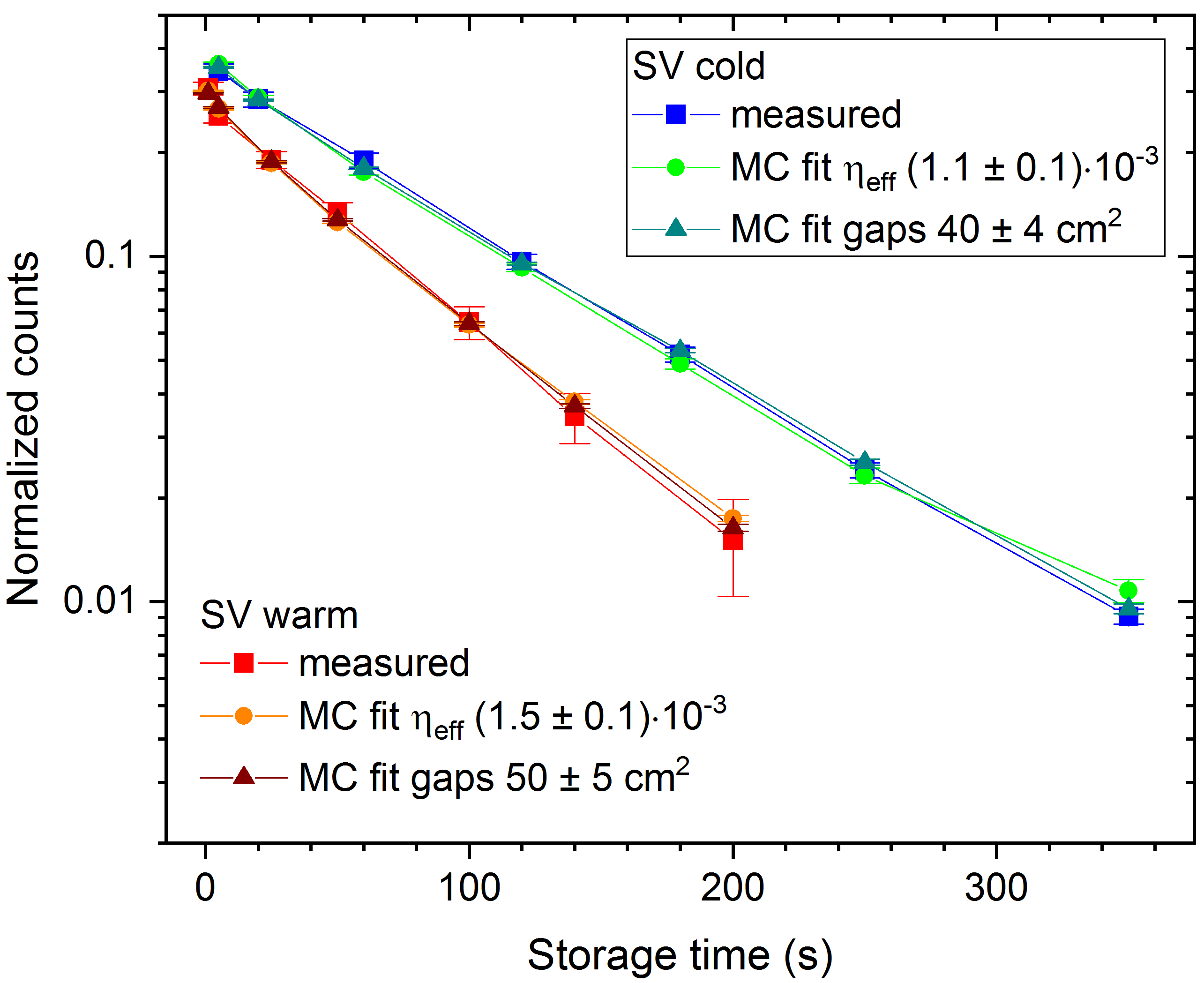}
}
\caption [MC of storage time] {
Simulation of the storage curve of the UCN source volume
for the cold and warm states. First we determined the effective loss parameter $\eta_\text{eff}$ by setting no gaps (orange and green circles, respectively). Then the calculations were redone by setting  $\eta$ to the literature value~\cite{Atchison2005c}, see text, and scanning the amount of gaps. A total gap of 50\,cm$^2$ corresponds to a surface fraction of 5$\times$10$^{-4}$. The filled triangles show the results of simulations with the $\eta$ from~\cite{Atchison2005c} and the best fit gap sizes.
}
\label{fig:MC-storage-time}
\end{center}
\end{figure}

The parameters of the storage vessel of the UCN source were obtained 
from matching MC results to the storage curves measured via 
leakage counts (see Sec.~\ref{sec:emptying}  
and Fig.~\ref{fig:UCN-West-1}).
Figure~\ref{fig:MC-storage-time} compares MC results with the measurements.

With all shutters closed and setting no gaps in the geometry, 
the dominating 'loss' channel is 
represented by the loss parameter within this vessel. 
In case if possible gaps should be co-represented by this parameter, 
we call the latter `effective'.
The MC scan gave a fit value
for the effective loss parameter
of (1.5$\pm$0.1)$\times$10$^{-3}$
for the vessel at room temperature, 
and (1.1$\pm$0.1)$\times$10$^{-3}$ 
for the vessel in the cold operating state (around 80\,K).

To interpret these large fit values, one must also consider that the effective loss 
parameter used in the simulation must include all the loss channels possible in the real system.  
These are (i) nuclear absorption and up-scattering, 
(ii) losses through gaps along the surface boundaries, and 
(iii) losses through holes distributed throughout the coated surfaces. 

Regarding (i), previous measurements at room temperature~\cite{Atchison2005c}
yielded an average loss parameter $\eta$ for DLC on 
aluminum (4$\pm$0.2)$\times$10$^{-4}$.   
At 70\,K the measured loss parameter in~\cite{Atchison2005c} 
was (1.7$\pm$0.1)$\times$10$^{-4}$. 

Regard (ii), 
gaps were found along the borders of individual 
surfaces (flaps, shutters, wall sections). 
and measured in the construction phase of the UCN source vessel.  
About a total of 25\,cm$^2$ were estimated for the bottom of the vessel, 
and about the same amount on the side walls. 
This results in a total gap size of 50\,cm$^2$ 
which is a fraction of 5$\times$10$^{-4}$ from 
all DLC-coated surfaces in the storage vessel. 
As this was measured at room temperature a large uncertainty might be 
on this value when comparing to operating conditions. 

\begin{figure}[htb]
\begin{center}
\resizebox{0.49\textwidth}{!}{\includegraphics{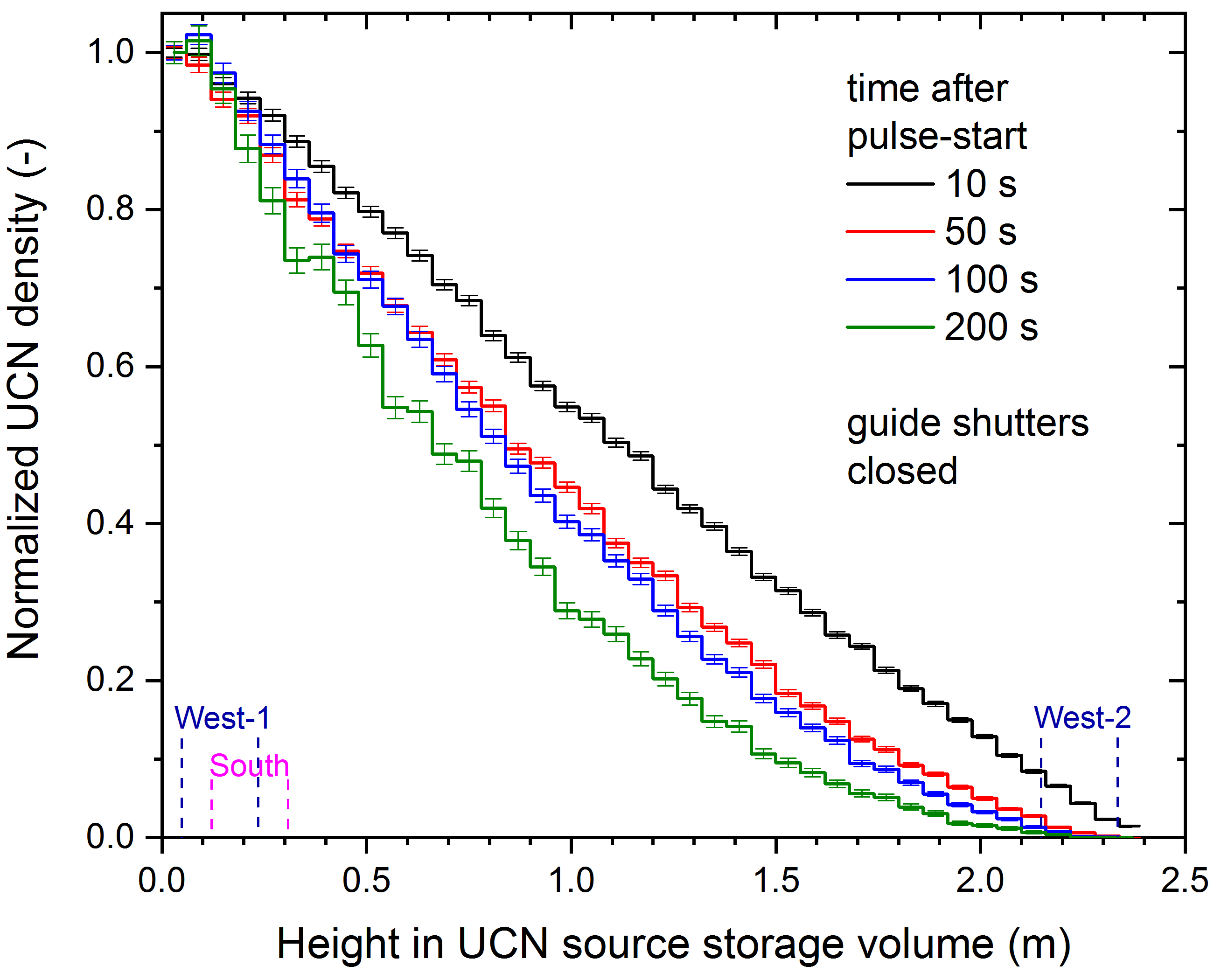}
}
\caption [Density vertical profile guide shutters closed] {
Simulated vertical distributions of the UCN density in the UCN source 
volume ($h_\text{max}$=2.4\,m) for various times after starting the production pulse. The latter in this example lasted 2 s with complete closure of the flapper valve 1 s after the end of the pulse.
The density profiles were normalized to the value at the bottom of the vessel. The vertical positions of the three UCN guide exits are indicated. The softening of the UCN spectrum with time is clearly visible.
At times close to the measured storage time constant of 90\,s 
in Fig.~\ref{fig:MC-storage-time} (see Tab.~\ref{table:STC-per-year}), 
the half-height density is about one third of the bottom-value. 
}
\label{fig:DensityVerticalProfileGuideShuttersClosed}
\end{center}
\end{figure}

It was important to check if this gap size is consistent with the measured UCN storage curves. Thus an additional series of simulations was performed.  The loss per bounce parameter was set to the measured literature value assuming a very similar surface quality as in~\cite{Atchison2005c}. Two gap regions were modeled: (i) an absorbing flat ring on the bottom of the volume around the main flaps, and (ii) an absorbing vertical stripe along one sidewall in the full height of the volume. 
These two positions represent in the simulation different aspects because 
of the gradient in UCN density in the storage vessel. 
In Fig.~\ref{fig:DensityVerticalProfileGuideShuttersClosed} we plotted density profiles 
for different times after the pulse start, 
also indicating the vertical positions of the beamlines. 
There is also a kinetic energy gradient along the vertical direction 
due to gravity causing a non-constant bounce-rate and loss probability. 
The absorbing `gap' surfaces were set variable. 
Since the mechanical measurements gave a similar surface of gaps for both the bottom and sidewalls, we kept this ratio during scanning this simulation parameter. 
We determined the most probable amount of gaps  by minimizing the $\chi^2$ between the measured and simulated data-points of the storage curve. 
With this model we also include possible holes in the DLC coating, the aforementioned third loss channel. 

The results with the best MC fits were included in Fig.~\ref{fig:MC-storage-time}. The uncertainty was estimated to 10\%. The fitted gap parameter matches very well the gaps determined during the construction i.e. a total of 50$\pm$5\,cm$^2$ for the warm state. In the cold state of the storage vessel the total gap parameter decreased to 40$\pm$4\,cm$^2$, however, remaining compatible with the room-temperature value within the errors. Thermal contraction of aluminum surfaces from 300\,K to 100\,K would yield only a 1\% change, however, depending on the initial size of a gap between the surfaces, this could change by a much larger percentage.

If using a harder UCN energy spectrum than our linear reference spectrum emerging from the UCN converter in the simulations, 
the fitted loss parameters would shift to slightly lower values. 
We will study this aspect in the future, after calibrating the initial energy spectrum 
by a combination of further test measurements and simulations. 
It is also important to note that a variation of the fraction of diffuse reflections 
in the storage volume did not make any difference at our level of accuracy. 

To conclude, the MC storage curves shown in Fig.~\ref{fig:MC-storage-time}
well match the ones measured at room temperature
and in the cold state (Fig.~\ref{fig:storage-classic-cold-RT}).
The simulated storage time constants agree with the measurement within the error bars.
Consequently, the linear energy spectrum assumed for UCN exiting from the sD$_2$ converter 
was a sufficiently good approximation for this purpose.

\subsection{UCN guide parameters from emptying time constants}

\begin{figure}[htb]
\begin{center}
\resizebox{0.49\textwidth}{!}{\includegraphics{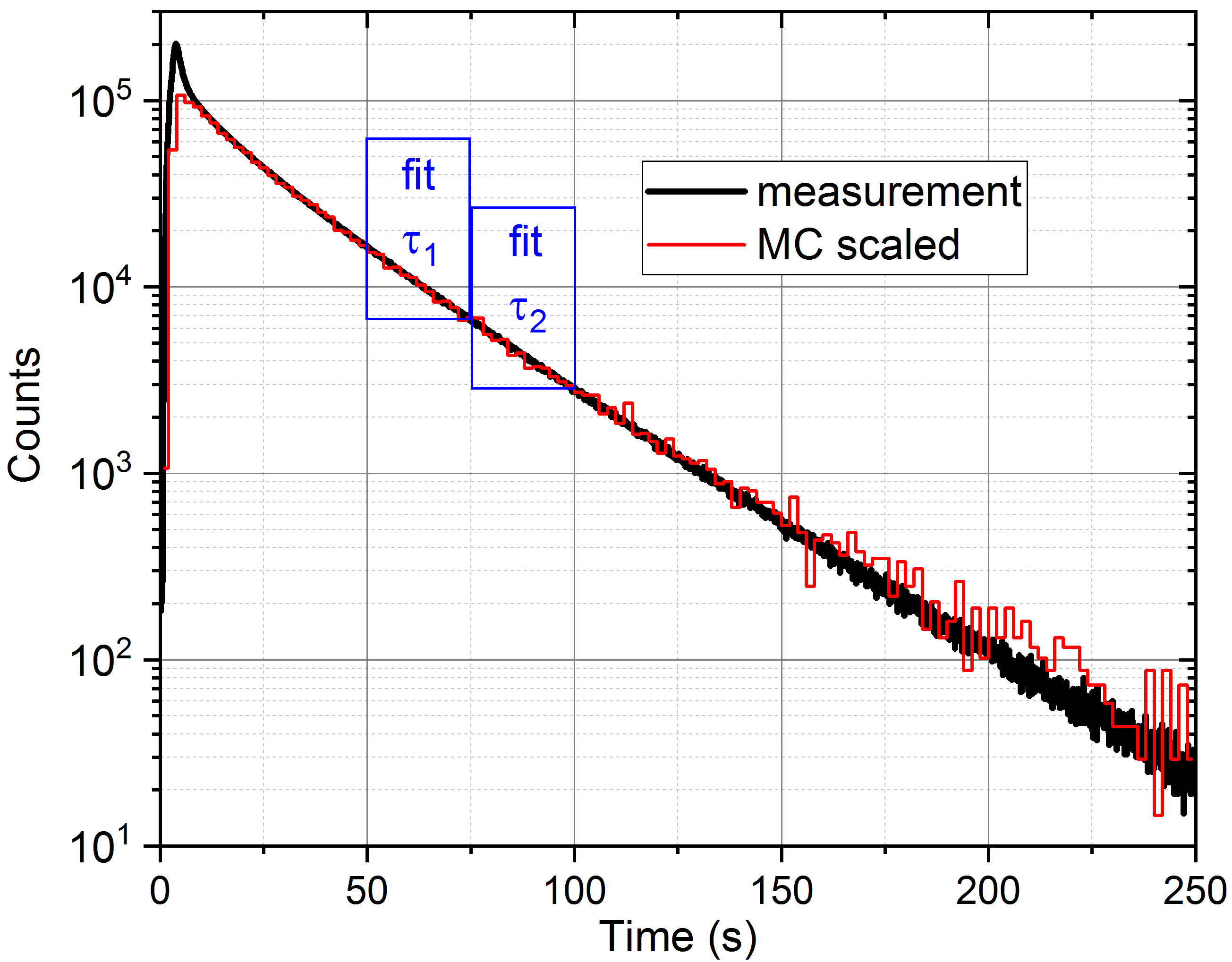}
}
\caption [MC of emptying profile] {
Comparison of emptying curves from MC simulations and measured data 
observed after a 3\,s proton beam pulse on UCN guide South. 
Two time intervals were defined as indicated in the figure and explained in the text. To each interval belongs one time constant, $\tau_1$ or $\tau_2$, from a single-exponential fit. The MC counts were scaled with a free fit parameter to best match the measured counts in the two intervals.
}
\label{fig:MC-emptying-time}
\end{center}
\end{figure}

\begin{figure}[htb]
\begin{center}
\resizebox{0.49\textwidth}{!}{\includegraphics{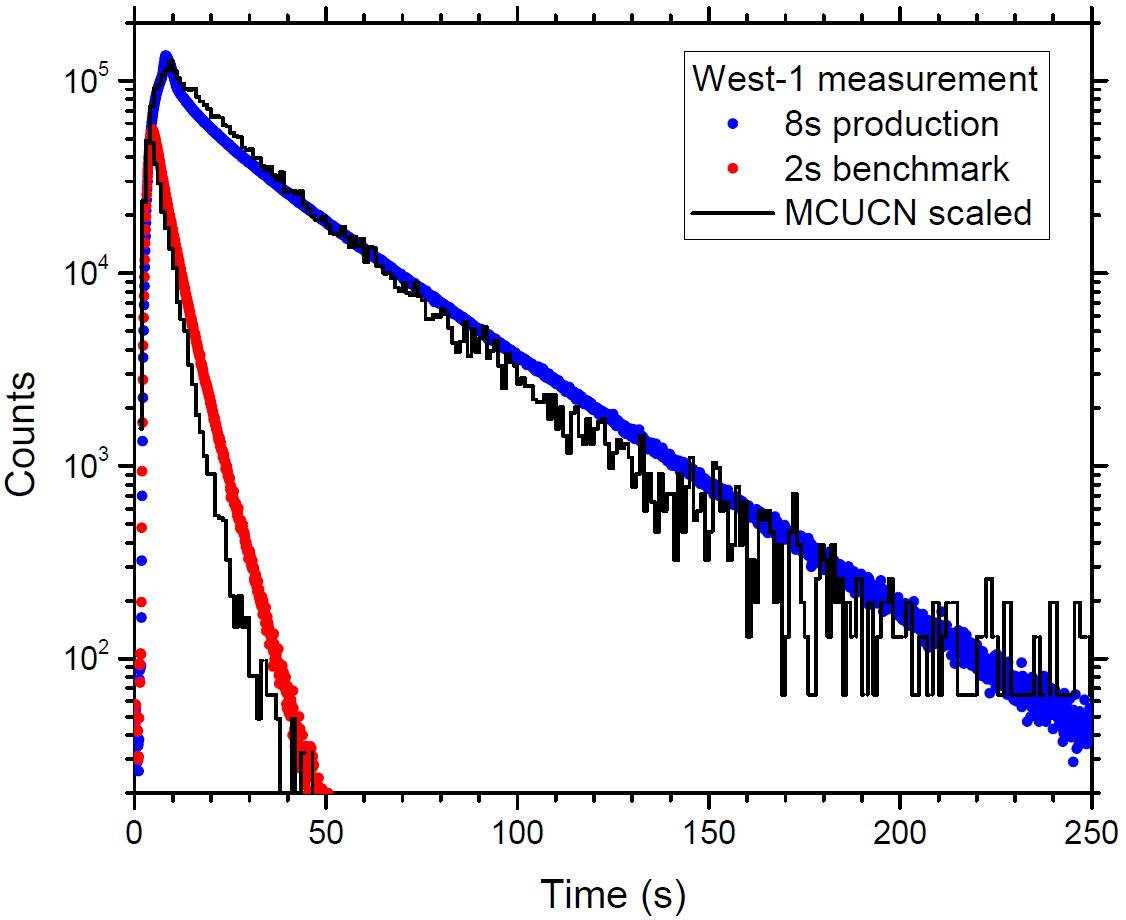}
}
\caption [MC of emptying curve West-1] {
Comparison of emptying curve observed on the West-1 beamport
in benchmark pulses and production pulses. The MC counts were scaled manually with the same value for both curves to fit the experimental curve of the 8\,s pulse in the interval after 50\,s similar to the case shown in Figs.~\ref{fig:MC-emptying-time}. 
}
\label{fig:MC-West1-Pulse}
\end{center}
\end{figure}

\begin{figure}[htb]
\begin{center}
\resizebox{0.49\textwidth}{!}{\includegraphics{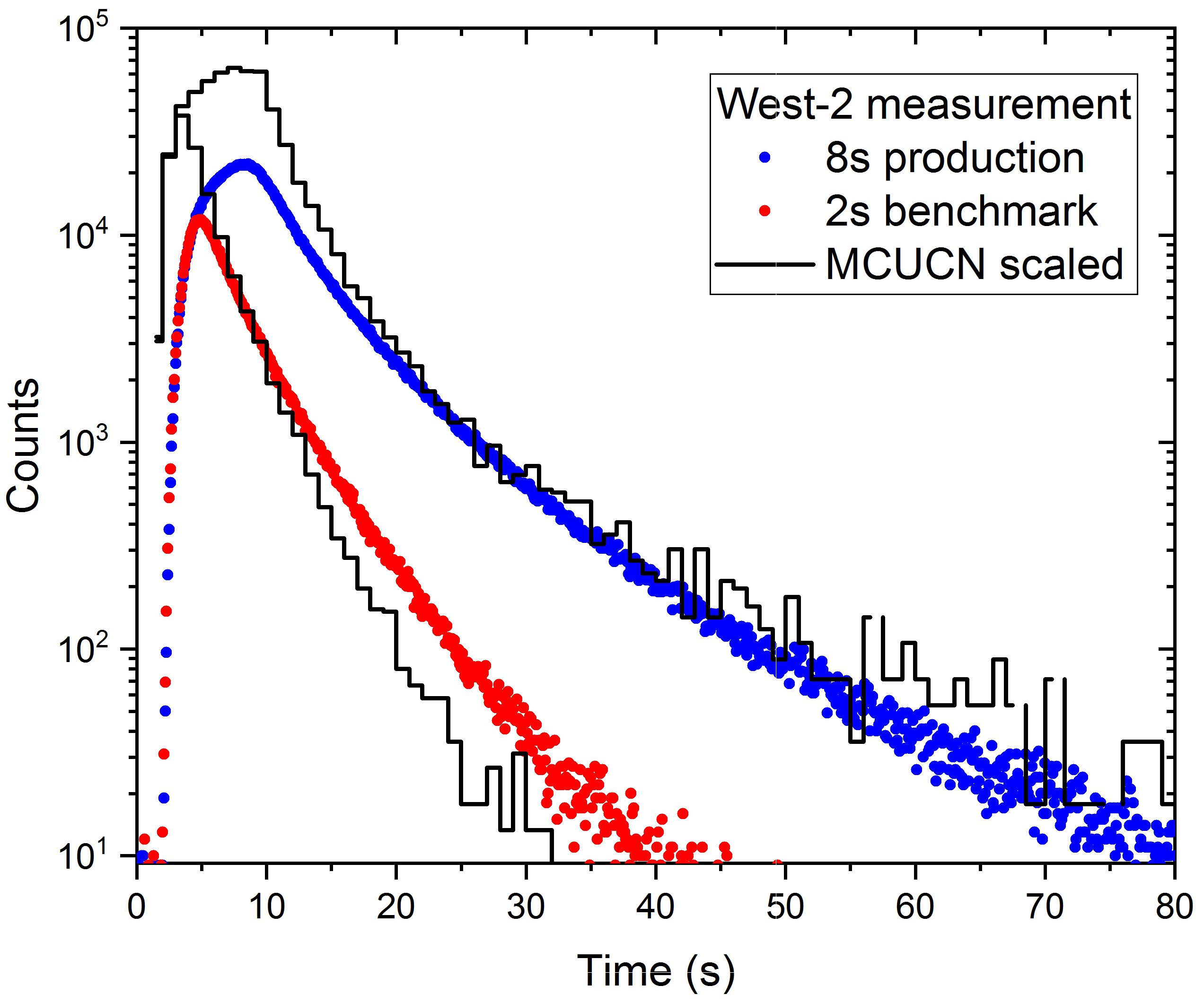}
}
\caption [MC of emptying curve West-2] {
Comparison of emptying curves observed on the West-2 beamport
in benchmark pulses and production pulses. The MC counts were scaled arbitrarily and independently for the 2 and 8\,s pulses in order to compare the slopes visually. 
}
\label{fig:MC-West2-Pulse}
\end{center}
\end{figure}

In Figs.~\ref{fig:MC-emptying-time},~\ref{fig:MC-West1-Pulse} and~\ref{fig:MC-West2-Pulse}
we compare measurement results and MC data.
The emptying curve displays the UCN counts decreasing 
in time after a benchmark pulse or a production pulse.
For the case of the South beamline, shown in Fig.~\ref{fig:MC-emptying-time}, the simulation was already validated via parameter scans and fit to the measured slopes. In Fig.~\ref{fig:MC-West1-Pulse}, the small mismatch in slope  between simulation and measurement indicates that the coating parameters obtained for beamline South are not exactly the same as for West-1.

We see a mismatch between MC and measured data at times comparable 
with the length of the initial proton beam pulse. The agreement is much better at later times
when, as discussed above, we expect that the details of the diffuse reflection model is less relevant. 
During the pulse it is likely that one measures a 
higher background which is not simulated, mostly from faster (not storable) neutrons
which can pass with low probability at small grazing reflection angles along the guides.

Another cause of the discrepancy at such short times could be 
that in reality the low roughness and high roughness surface 
regions (the latter including cavities at the edges) 
are unevenly distributed, whereas in the MC model we set uniform surface quality. 
At short time-scales and, in contrast to what was previously discussed for equilibrium conditions, the 
implementation of diffuse reflections in the model actually matters. 
An additional simplification was that the assumed energy spectrum emerging from the UCN converter was linear as obtained from simulations of an optimal sD$_2$ structure. If a harder UCN spectrum would be generated, it could enhance the higher energies at early times.

\begin{figure}[htb]
\begin{center}
\resizebox{0.49\textwidth}{!}{\includegraphics{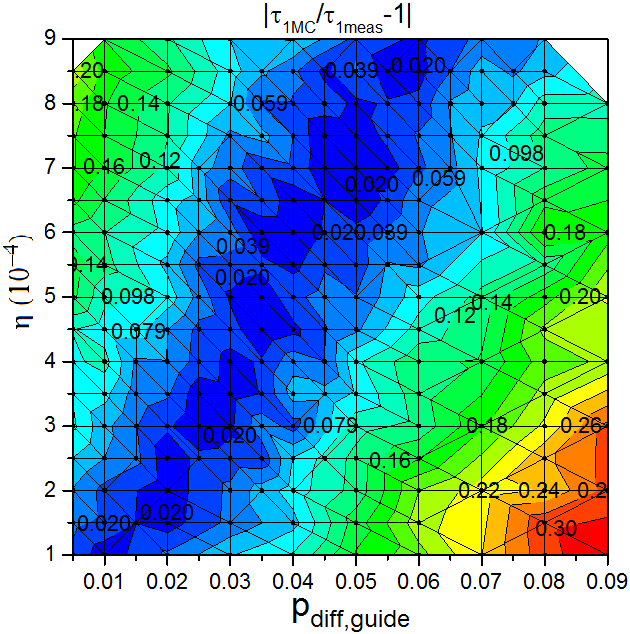}
}
\caption [MC of emptying tau1] {
Relative deviation plot in the $\eta-p_\text{diff,guide}$ plane between the MC and the experimental 
emptying time constant $\tau_1$ (see text). The colors emphasize the contour lines.
}
\label{fig:MC-emptying-tau1}
\end{center}
\end{figure}

\begin{figure}[htb]
\begin{center}
\resizebox{0.49\textwidth}{!}{\includegraphics{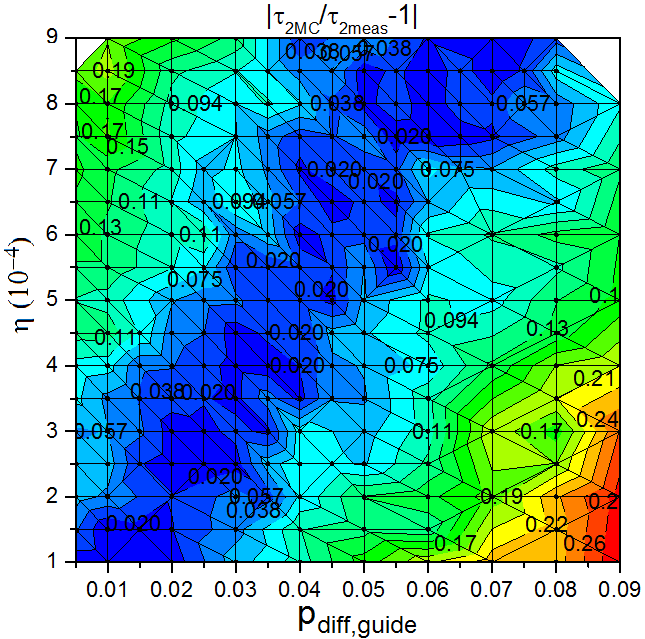}
}
\caption [MC of emptying tau2] {
Relative deviation plot in the $\eta-p_\text{diff,guide}$ plane between the MC and the experimental emptying time constant $\tau_2$ (see text).  The colors emphasize the contour lines.
}
\label{fig:MC-emptying-tau2}
\end{center}
\end{figure}

In order to benchmark MC simulations with measurements we compare the emptying 
time constants which do not depend on the scaling of the UCN counts. 
We defined two separate time intervals as indicated by the rectangles along 
the data in Fig.~\ref{fig:MC-emptying-time}, one
from 50\,s to 75\,s, and the next from 75\,s to 100\,s. 
Within these short intervals we can safely assume single-exponential dependence 
and sufficient statistics for the benchmark. 
To each interval corresponds 
one constant, $\tau_1$ or $\tau_2$.

Each of these two time constants were extracted from the measured 
and the simulated time spectra. 
We calculated the relative deviation between the MC and 
the experimental value. 
The scope is to scan the parameter plane, one axis being the loss parameter $\eta$ 
and the other axis the diffuse reflection fraction $p_\text{diff,guide}$. 
A single simulation takes ten minutes in order to produce enough statistics 
(goal: below 1.5\% relative error) for only one parameter configuration. Therefore 
we scanned along a uniform grid in the parameter space and and used triangular interpolation. 
Thus we obtained a continuous surface of the distribution of the absolute 
value of the relative deviation on the third axis. 
Along the third axis we can define contour levels of the deviation.

We interpret the 1$\sigma$ of the measurement and also of the simulation 
as 68\% confidence interval,
assuming that the convolution of several uncertainty contributions 
result in a close-to-normal distribution. 
Since we have statistical spread on both the measured and the 
MC values, 1$\sigma$ here is the square root of the sum of squares of 
each 1$\sigma$ uncertainty.

Figures~\ref{fig:MC-emptying-tau1} and~\ref{fig:MC-emptying-tau2}
show the MC results. 
These reveal a straight valley with a lowest deviation between MC and 
measurement reaching down to about 2\% 
with almost connected islands. 
We take as reference the next level, 4\%, 
which is the first fully connected area. 
This corresponds to about 2-3$\sigma$ deviation. 
In the calculations we had a 2\% relative statistical spread in the MC counts, and in case of the measurements below 0.4\% of the counts.

We observe that the valley of lowest deviation points into a direction 
in which both parameters grow simultaneously. 
This is because, in this case, the UCN experience two strongly competing effects: 
(i) the emptying time-constants are on one hand determined by the speed of emptying, i.e. the transmission quality, 
but (ii) on the other hand by the loss time constant of the entire 
storage vessel and neutron guide system. 
For example, if we increase the loss parameter, 
the time constant will decrease because we lose UCN much faster.
At the same time, if we increase the diffuse parameter 
we make the emptying slower and thus compensate a faster decay 
in the time spectrum.

\subsection{Consistency check by varying the storage volume diffuse fraction}

Since the simulations are limited by computation time, 
we cannot perform very detailed scans by adding those parameters 
to the parameter space for which we concluded that the dependency 
of the fit to experiment must be small. 
However, as a consistency check, we still performed several 
calculations in which we changed the values of such less relevant parameters.

\begin{figure}[htb]
\begin{center}
\resizebox{0.49\textwidth}{!}{\includegraphics{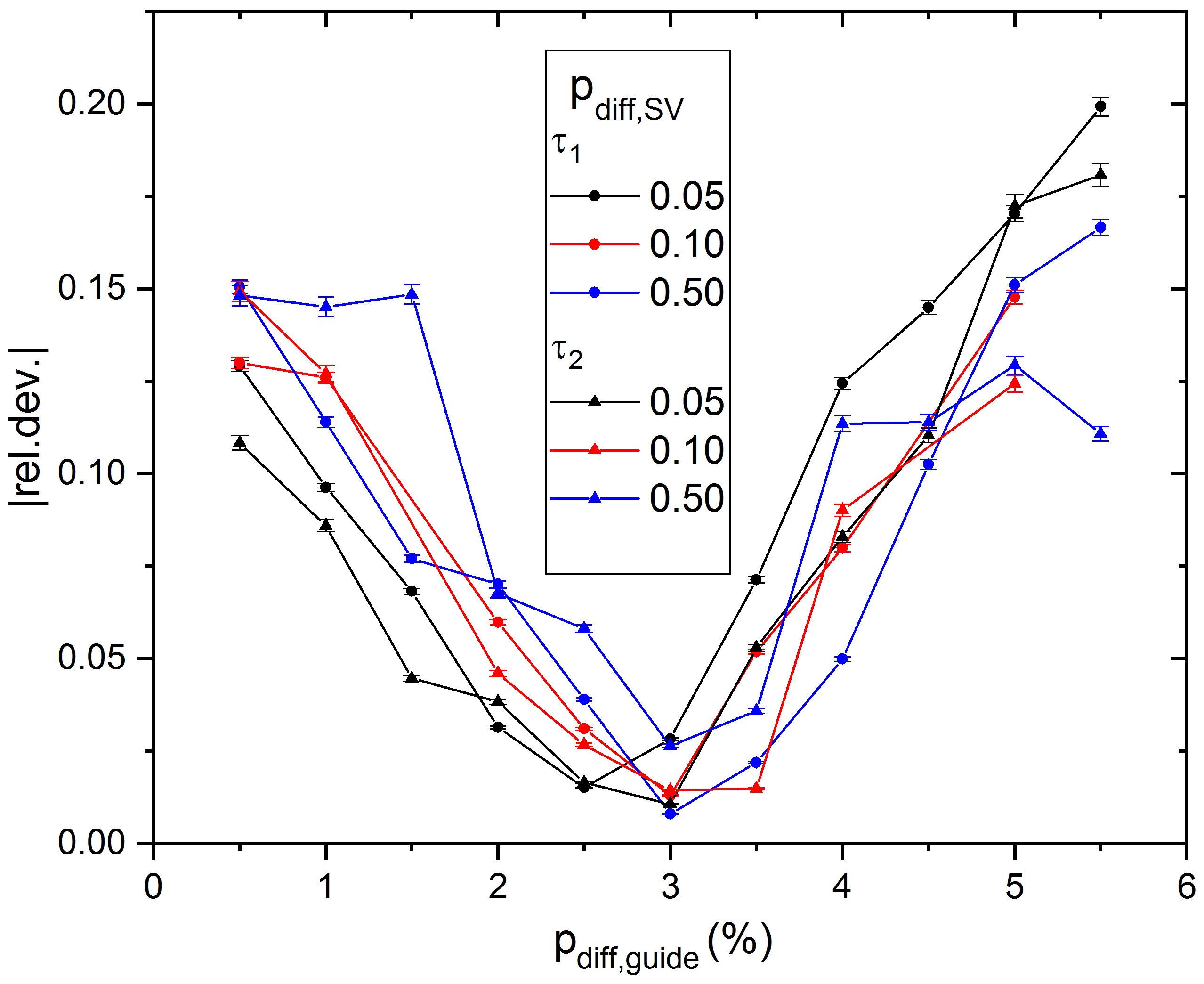}
}
\caption [MC tau versus $p_\text{diff,guide}$ and $p_\text{diff,SV}$] {
Relative deviation of time-constants as a function of the fraction of diffuse reflections in the storage volume, $p_\text{diff,SV}$. The scan was performed  across the $\eta$ and $p_\text{diff,guide}$ parameter plane of the neutron guides.
}
\label{fig:MC-tau-versus-pdiff}
\end{center}
\end{figure}

In Fig.~\ref{fig:MC-tau-versus-pdiff}
we show a scan to check the variation of the two 
aforementioned emptying time-constants as a 
function of the fraction of diffuse reflections at the storage vessel surface. 
The MC scan was done along a line in the $\eta-p_\text{diff,guide}$ plane across the minimum deviation valley visible in 
Figs.~\ref{fig:MC-emptying-tau1} and~\ref{fig:MC-emptying-tau2}.  Different values for the fraction of diffuse reflections  
in the storage volume were set. 
We see a weak dependence of the minimum well on this parameter, however, within the uncertainty of the extracted neutron guide parameters. This confirms the robustness of our results with respect to this parameter.

\subsection{UCN intensity variation from the magnetic field of the polarizer}
\label{sec:mag-simulation}

The measurements described in Sec.~\ref{sec:magnet} were 
reproduced satisfactorily by simulations as illustrated in Fig.~\ref{fig:totalUCNvsB}.
The low-energy range is not yet well modeled, either because of the not sufficiently well known low-energy spectrum or the not well known transmission of the low-energy UCN.
The SC polarizer magnet was simplified to a rectangular potential 
profile along the beam axis located at the position of the safety window. 
Further studies are planned. 
We also envisage to check how the alternative diffuse reflection models would affect the results.

\begin{figure}[htb]
\begin{center}
\resizebox{0.49\textwidth}{!}{\includegraphics{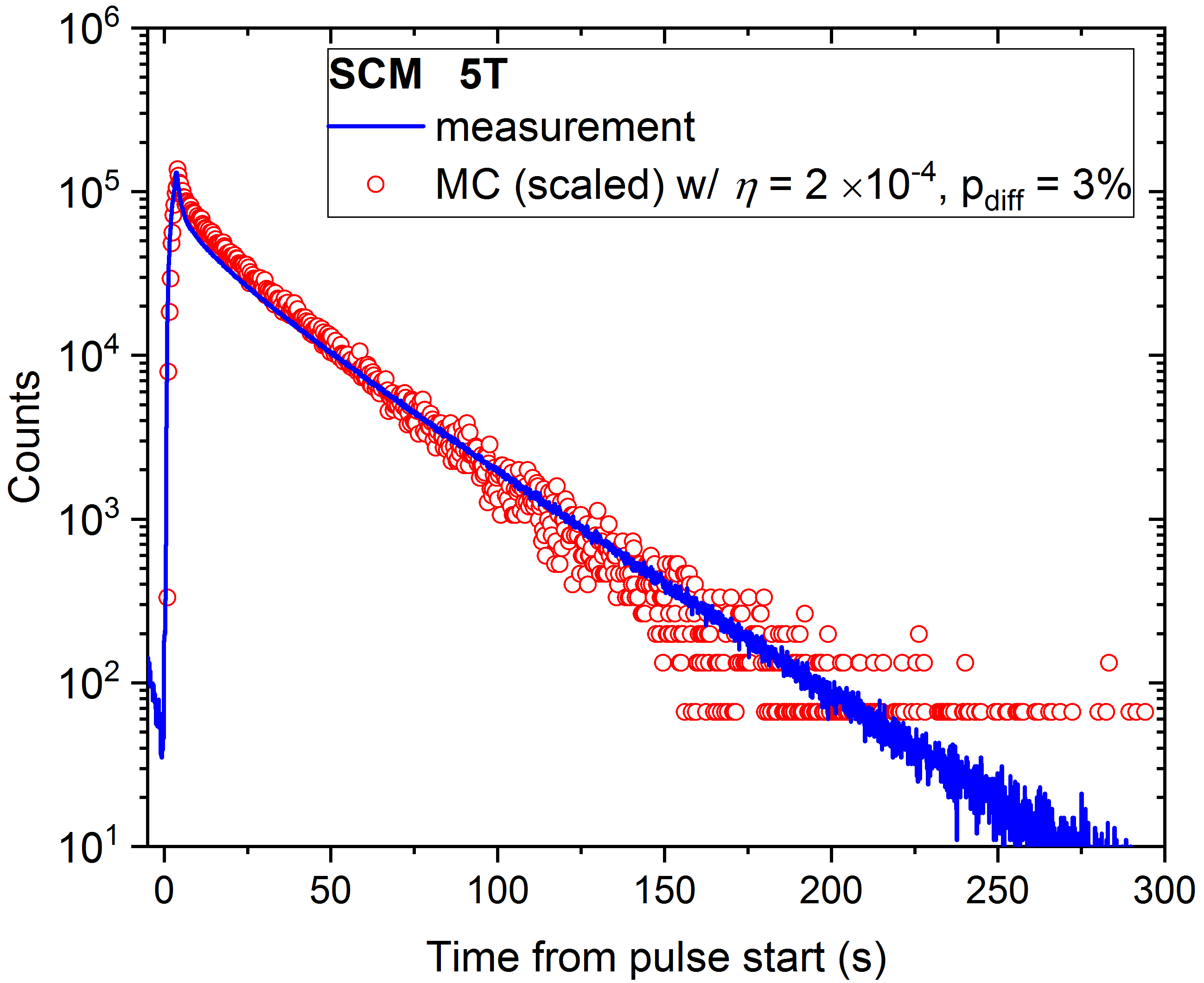}
}
\caption [CompareMeasMCUCNwithSCMat5T-linear] {
Comparison of the measured and simulated emptying curves at 5\,T field in the SC magnet. 
}
\label{fig:CompareMeasMCUCNwithSCMat5T-linear}
\end{center}
\end{figure}

\begin{figure}[htb]
\begin{center}
\resizebox{0.49\textwidth}{!}{\includegraphics{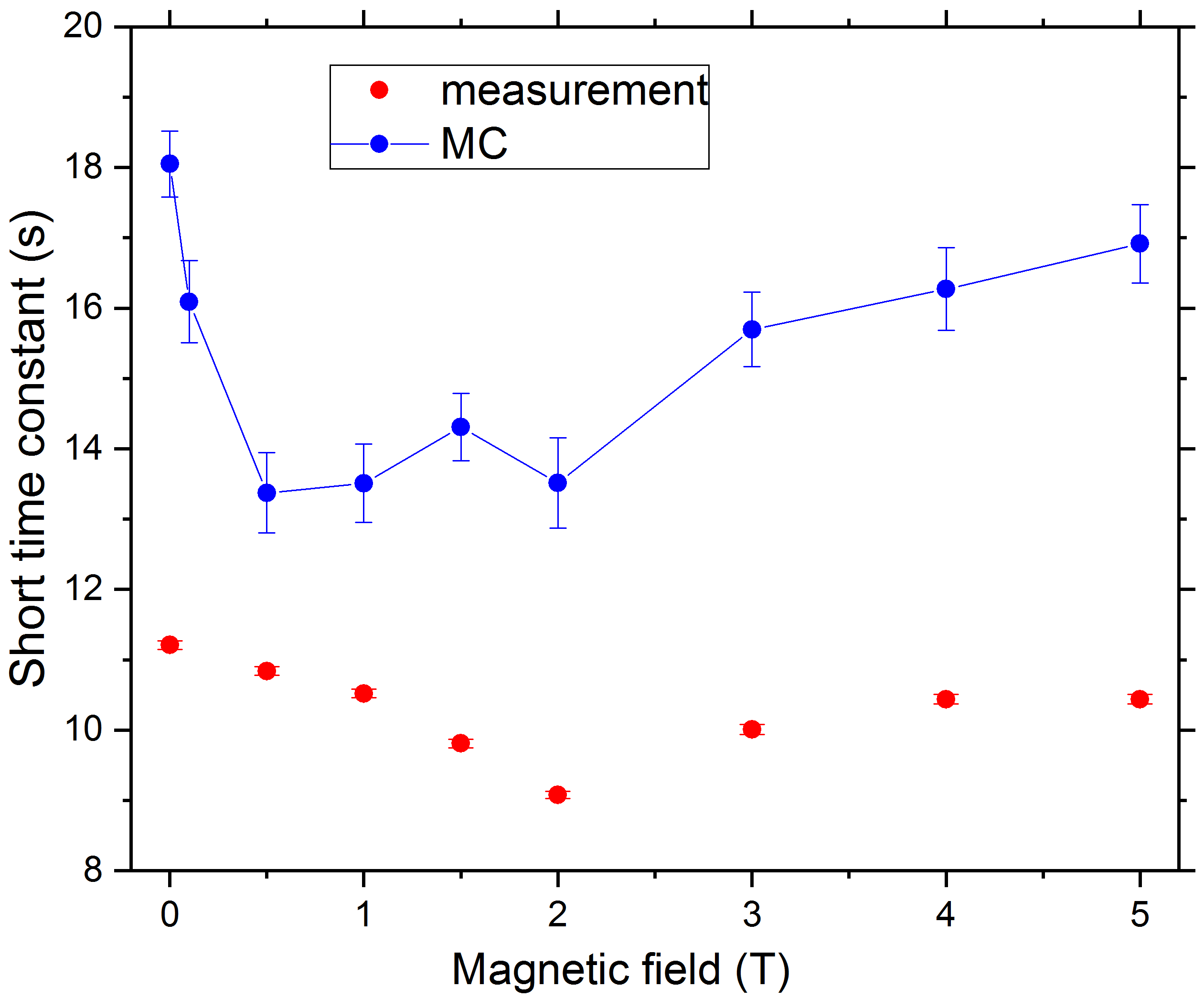}
}
\caption [MCUCN-short-time-constant-vs-SCM] {
MC simulation results and the measured short 
time constant (Fig.~\ref{fig:EmptyingTimevsB})
during emptying 
for a given magnetic field in the SC magnet. 
}
\label{fig:MCUCN-short-time-constant-vs-SCM}
\end{center}
\end{figure}

In Fig.~\ref{fig:CompareMeasMCUCNwithSCMat5T-linear} the emptying curves 
for the case of a 5\,T magnetic field 
 are depicted. 
It can be observed that for time-scales below 50\,s the simulation does not reproduce very well the measurement. 
One cause, which will be checked in further studies, could be that the short time characteristics are determined by the concrete modeling of the diffuse reflections. A Lambert-type modeling turns out to be a too rough approximation for short detection times. It is sufficient only for detection times larger than several tens of seconds as demonstrated in Figs.~\ref{fig:MC-emptying-time} and~\ref{fig:CompareMeasMCUCNwithSCMat5T-linear}.
The simulations overestimate the time constant 
for short times as visible in Fig.~\ref{fig:MCUCN-short-time-constant-vs-SCM}. 
However, the profile indicating a minimum was qualitatively reproduced. 
This can be due to the competition of two effects both increasing with the magnetic field: (i) better transmission through the aluminum window allowing for shorter detection times, and (ii) more UCN rejected by the polarizer which depolarize again and have a second chance later to be transmitted, and thus increasing the detection time.

\subsection{Simulations of the UCN transmission from the solid deuterium to the beamport}

In this section we provide details from simulations describing the dependence of important quantities on time, position, energy,
which are not accessible directly from experiments for practical reasons. However, they 
provide important information on the characteristics of the PSI UCN source and UCN guide system. 
 We obtain information on, for example (i) the time-integrated transmission between the upper surface of the UCN converter and 
the end of the UCN guides, including the detector window, or (ii) on the energy spectra as a function of time of the detected UCN. 
Another not accessible observable, the density profile of UCN 
along the vertical direction in the storage vessel of the UCN source 
was already plotted in Fig.~\ref{fig:DensityVerticalProfileGuideShuttersClosed}, 
and helped to estimate the ratio of bounce-rates at different heights.

The parameters for coating quality in the simulation results presented next are summarized in Table~\ref{table:SimulationParameters}. We considered the case of direct transmission between the lid above the sD$_2$ and the detector window, including these two.

\begin{table*}
\centering
\begin{tabular}{|c|c c|c c|c c|c c|}
  \hline
Surface  & $V_\text{F}$ & Method &  $\eta$  & Method &  $p_\text{diff}$ & Method &  $\Sigma_\text{atten}$ & Method \\
            & (neV)  &   &  (-)   &   &  (-)  &   &  cm$^{-1}$m/s  &   \\
	\hline
Lid sD$_2$  & 54  & Al calc.  &  1$\times$10$^{-4}$   & low dep. &  0.0  & low dep.  &  49  & meas. \\
	\hline
Vert. guide  & 230  & NiMo calc. &  5$\times$10$^{-4}$   & gaps calc. &  0.04  & low dep. &  $\infty$  & design \\
	\hline
Stor. vessel  & 230  & DLC meas. &  11$\times$10$^{-4}$   & meas.+MC &  0.10  & low dep. &  $\infty$  & design \\
	\hline
Guide South  & 220  & NiMo meas. &  3$\times$10$^{-4}$   & meas.~\cite{Bondar2017} + gaps &  0.02  & meas.+MC &  $\infty$  & design \\
	\hline
Windows  & 54  & Al calc.  &  1$\times$10$^{-4}$   & low dep. &  0.0  &  low dep.  &  77  & meas. \\
	\hline
\end{tabular}
\caption[Simulation parameters]{Parameters of the coatings used in the simulations: optical potential ($V_\text{F}$), loss parameter ($\eta=W/V_\text{F}$), fraction of diffuse (Lambert) reflections ($p_\text{diff}$), and attenuation constant of the material ($\Sigma_\text{atten}$) for 1\,m/s. These numbers represent (i) theoretical values~\cite{Golub1991}, (ii) measurements discussed in this paper and in~\cite{Bondar2017}, (iii) simulations benchmarked with measurements, (iv) geometrical estimations of gaps, and (v) rough estimations with a low dependency of the outcome.
}
\label{table:SimulationParameters}
\end{table*}

\begin{figure}[htb]
\begin{center}
\resizebox{0.49\textwidth}{!}{\includegraphics{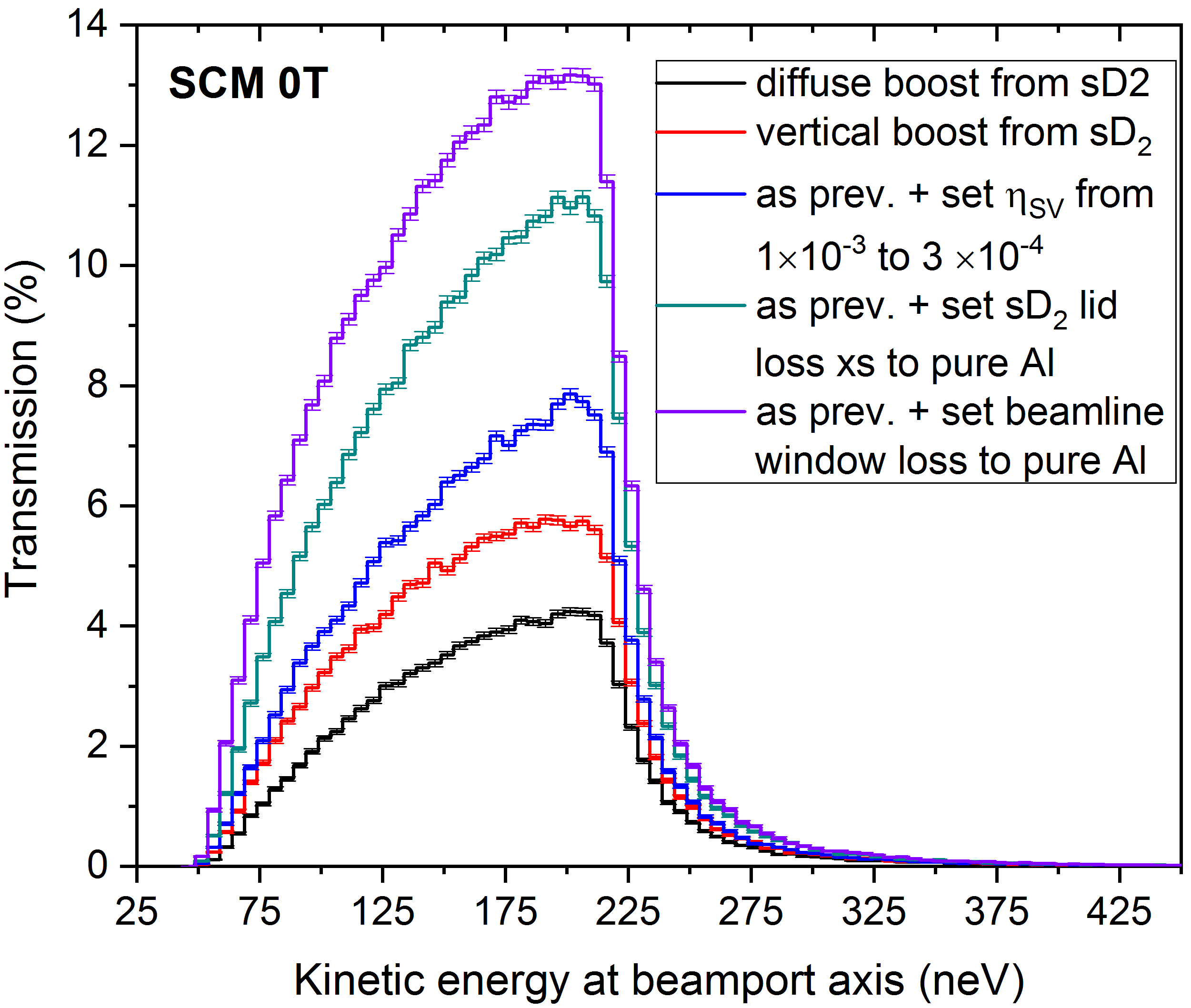}
}
\caption [Transmission sD2 to beamport 0T] {
Transmission between the UCN converter and detector 
(including the Al foil of the detector) when the 
polarizer is switched off. 
We counted all UCN having the chance to exit the beamport until the next pulse.
The kinetic energy is given with respect to the height of the beamport axis.
}
\label{fig:TransmissionSD2toBeamlineExit-SCM00T}
\end{center}
\end{figure}

\begin{figure}[htb]
\begin{center}
\resizebox{0.49\textwidth}{!}{\includegraphics{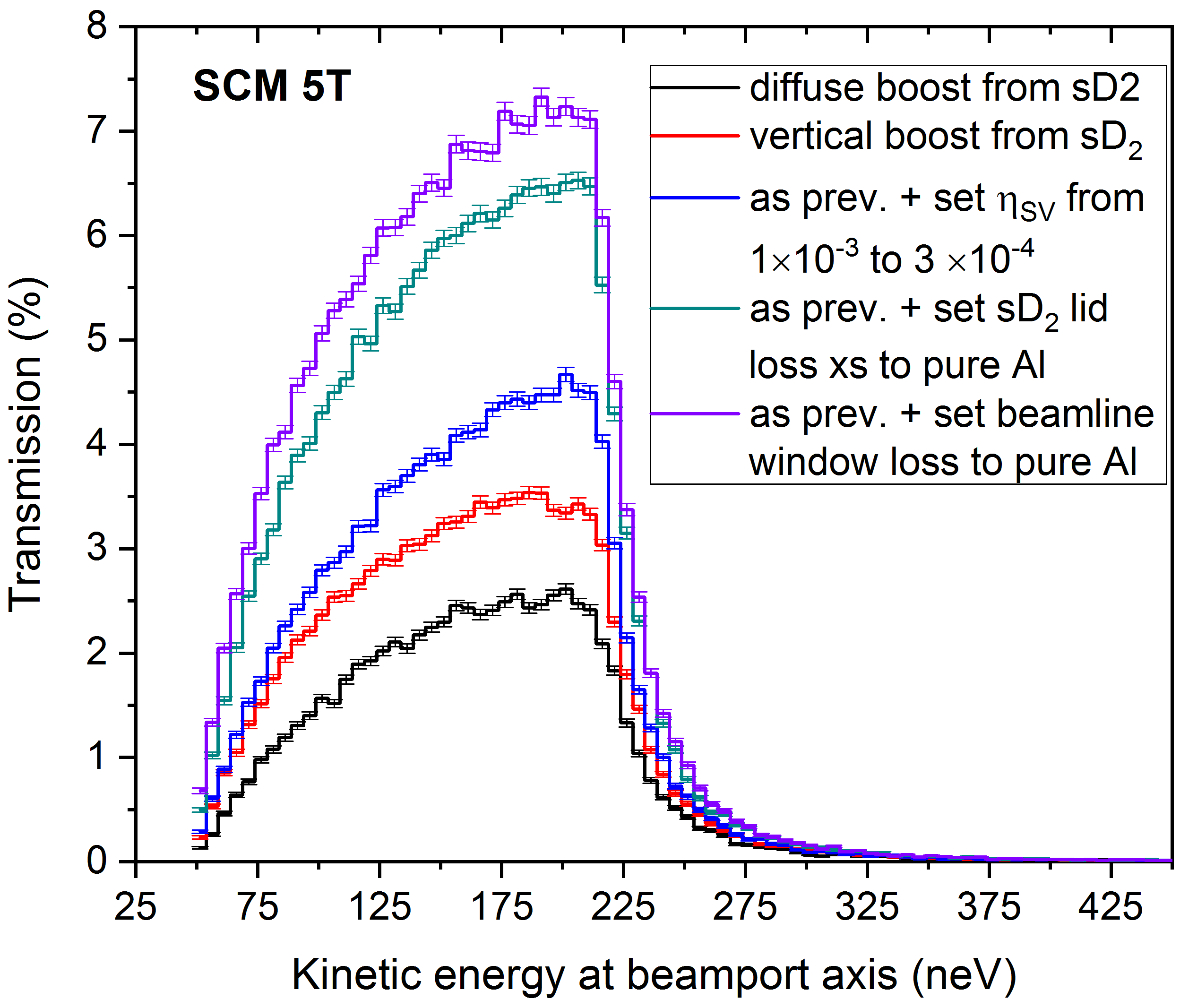}
}
\caption [Transmission sD2 to beamport 5T] {
Similar as in Fig.\ref{fig:TransmissionSD2toBeamlineExit-SCM00T}, 
however, with the SC polarizer at 5\,T magnetic field.
}
\label{fig:TransmissionSD2toBeamlineExit-SCM05T}
\end{center}
\end{figure}

In Figs.~\ref{fig:TransmissionSD2toBeamlineExit-SCM00T} and~\ref{fig:TransmissionSD2toBeamlineExit-SCM05T} the transmission of UCN between the solid deuterium surface and the detector is shown for zero and for 5\,T field in the SC polarizer magnet. The transmission was calculated as the ratio of the UCN counts after the detector window, and the number of UCN generated below the lid of the converter vessel, and was separated into bins of the kinetic energy at the level of the beamline axis.
The vacuum window of the beamport and that of the detector set the lower energy boundary of the spectrum to the optical potential of aluminum.
The different lines indicate successive simulations in which we changed only one parameter at a time (except the random seed). 
For obvious reasons the transmission of UCN is strongly energy dependent, thus for a more general purpose, 
the transmission has to be calculated as a function of energy.

In a first step (see the lowest transmission curve in Figs.~\ref{fig:TransmissionSD2toBeamlineExit-SCM00T} and~\ref{fig:TransmissionSD2toBeamlineExit-SCM05T}) the UCN source 
and guide optics was set as previously benchmarked by matching the test measurements 
- as described in the above sections - and by setting the energy boost from the solid deuterium 
diffusely (see Sec.~\ref{sec:InitialTraj}). 
This kind of diffuse boost corresponds to the most probable case when the 
upper surface of the converter material is very rough, e.g. being composed by poly-crystalline structures, as frost~\cite{Anghel2018}. 

\begin{figure}[htb]
\begin{center}
\resizebox{0.49\textwidth}{!}{\includegraphics{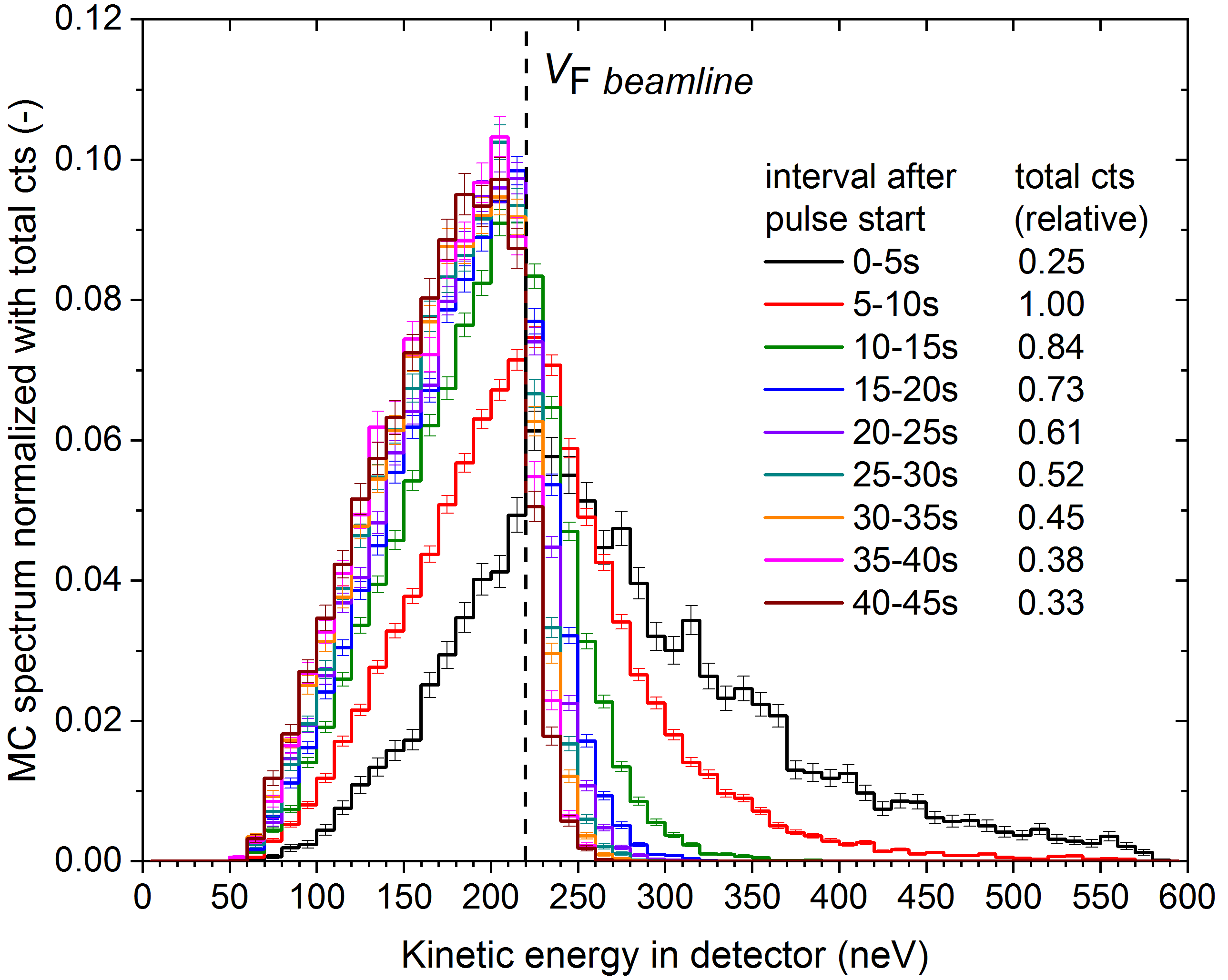}
}
\caption [Energy spectrum - West-1] {
UCN counts versus kinetic energy as simulated for a detector at the West-1 beamport.
The energy spectra of unpolarized UCN are calculated for different time intervals 
after the start of the 4\,s long proton beam pulse.
}
\label{fig:EnergySpectrumEvolutionWest-1DirectTransmission}
\end{center}
\end{figure}

The next case is the simulation of a vertical boost which would correspond to a perfectly flat 
surface of the converter material. 
The gain factor due to this new setting turns out to be 
about 1.4$\pm$0.1. 
The latter and next gain factors from these simulations were calculated by averaging in the UCN energy interval 120-230\,neV, which is a representative range in which most UCN are transmitted to the beamport (see transmission curves in Fig.~\ref{fig:EnergySpectrumEvolutionWest-1DirectTransmission}).

In a second step, we estimate the loss effect due to the gaps in the storage-vessel of the 
UCN source by changing the effective loss parameter from the one benchmarked 
above 1$\times$10$^{-3}$ to the pure DLC contribution 3$\times$10$^{-4}$ 
- as found in measurements~\cite{Atchison2005c}.   
The gain factor obtained in this step is 1.3$\pm$0.1.

The next curves help to localize the loss effects due to the absorption, 
up-scattering and back-scattering in the vacuum safety windows. 
It has been shown in~\cite{Atchison2009}  
and in later comparisons with
measurements reported in Section~\ref{sec:foil-transmission} and in Ref.~\cite{Ries2016}
that the loss cross-section in the aluminum window is 
a factor 2.2-2.5 larger than the theoretical value. 
At first we reduced only the loss cross section in the lid of the converter vessel 
to the theoretical one. 
The gain factor due to this new setting is 1.5$\pm$0.1. 

The last step  was switching the loss parameter in the vacuum safety window 
to the theoretical one. 
In this case we see a difference between the two magnetic field settings in the SC polarizer.  
The gain factor obtained in this step is 1.3$\pm$0.1 for field off (un-polarized UCN) 
and 1.1$\pm$0.1. for a 5\,T field (polarized UCN). 
The smaller difference for the case of polarized neutrons 
was expected because of the energy boost caused by the SC magnet.

The transmission profiles given as a function of UCN energy are valid independently from the initial energy spectrum which was generated at the level of the solid deuterium surface. 
One can estimate the energy spectra detected at the beamport assuming a linear initial spectrum. The calculations were again considering the case of direct transmission towards the detector.
The resulting energy spectra were plotted in Fig.~\ref{fig:EnergySpectrumEvolutionWest-1DirectTransmission} as a function of the time bins after the start of the 4\,s proton pulse. The spectra were normalized with their total counts. The change over time in the total counts (relative to the maximum) can be seen in the second column of the figure legend.
It can be seen that the spectrum is softening very fast just after the pulse and continues to do so with decreasing rate.
The dashed vertical line at 220\,neV indicates the optical potential of the UCN guide coating (NiMo) set in the MC simulation. 
One can clearly see the decrease in the spectra above this energy limit. 
At short times a large fraction of very-cold neutrons 
(in our case with energies higher than 220\,neV) are still present. 
Even after 30\,s the fraction of these faster neutrons is not negligible compared to the area of the same curve below the 220\,neV threshold. 

We can conclude that MC simulations help us to quantify several effects 
causing UCN losses in the PSI UCN source and guide system. 
Just after the proton pulse there is a considerable amount of 
neutrons faster than UCN, which could be important for specific applications.

\section{Summary}
\label{sec:summary}

This paper presents all components which define the 
ultracold neutron optics of the PSI UCN source.
UCN intensities and their increase in subsequent 
years are given.
Measurements of storage time constants, 
of UCN intensities and their time dependence,  
and of UCN transmissions are shown as relevant characterization measurements.
Together with measurements characterizing the UCN surfaces
they define the neutron optics parameters of the source.

The MCUCN simulation model
together with dedicated experiments allowed for constraining 
overall quality parameters characterizing the optical components 
above the main shutter of the UCN source vessel. 
A specific characterization of the vertical guide part of the source
will be the subject of a forthcoming study. 
The resulting cross-checked simulation allowed the calculation of relevant
parameters of the UCN source which are not accessible to experiments,
like the UCN density distribution in the storage vessel.
The presented results 
demonstrate the excellent performance of the PSI UCN source
and 
will help us to better understand 
and further optimize its performance.

\section*{Acknowledgments}

All people who have been contributing to design and construction of the 
PSI UCN source are gratefully acknowledged.
The outstanding continuous technical support of M.~Meier and F.~Burri 
is acknowledged.
We are grateful to the 
BSQ group operating the UCN source, namely
P.~Erisman, R.~Erne, K.~Geissmann, A.~Kalt, T.~Hofmann, J.~Welte and D.~Viol. 
Operations support by A.~Anghel is appreciated.
We acknowledge the excellent support by
the PSI proton accelerator operations section, especially A.~Mezger and D.~Reggiani,
and the target group of M.~Wohlmuther and V.~Talanov, and
the excellent support by many other PSI support groups.
Personal efforts and support of
P.~Bucher, U.~Bugmann, M.~D\"anzer, J.~Ehrat, A.~Fuchs, 
A.~Gn\"adinger, A.~Hirt,
M.~Horisberger, B.~Jehle, R.~K\"ach, R.~Knecht, M.~Koller,
M.~M\"ahr,
O.~Morath, M.~M\"uller, W.~Pfister, P.~R\"uttimann, 
R.~Schelldorfer, T.~Stapf and C.~Stettler
are acknowledged.
This work was supported by the Swiss National Science Foundation
Projects $200020\_137664$ and $200020\_149813$, $200020\_163413$. 
Special thanks for granting access to the PL-Grid~\cite{PLGrid}.

\bibliographystyle{unsrt}

\end{document}